\documentclass[%
    aps, 
    physrev,
    reprint,
]{revtex4-2}

\usepackage{amsmath}
\usepackage{amssymb}
\usepackage{bm}
\usepackage{braket}
\usepackage{comment}
\usepackage[normalem]{ulem}

\usepackage{printlen}
\usepackage{float}
\usepackage{graphicx}
\usepackage{xcolor}
\usepackage[caption=false]{subfig}

\usepackage{enumitem}

\usepackage{dcolumn}
\usepackage{multirow}
\usepackage{makecell}
\usepackage{tabularx}

\usepackage{tikz}

\usepackage[colorlinks=true, allcolors=blue]{hyperref}
\usepackage[capitalise]{cleveref}
\usepackage{orcidlink}

\newcommand{\fd}[1]{\mathsf{f}=#1}
\newcommand{\rot}[1]{\mathsf{Rot}(#1)}
\newcommand{\reg}[1]{\mathsf{Reg}(#1)}
\newcommand{\GHZ}[1]{\text{GHZ}(#1)}
\newcommand{\XY}{_\mathrm{XY}}
\newcommand{\fin}{_\mathrm{fin}}
\newcommand{\dat}{_\mathrm{dat}}
\newcommand{\anc}{_\mathrm{anc}}

\newcommand{\dg}{^\dagger}

\newcommand\Yale{
    Departments of Applied Physics and Physics, 
    Yale University, 
    New Haven, CT 06520, USA
    }
    
\newcommand\YQI{
    Yale Quantum Institute, 
    Yale University, 
    New Haven, CT 06511, USA
    }
    
\newcommand\IBM{
    IBM Quantum, 
    IBM T.~J. Watson Research Center, 
    Yorktown Heights, NY 10598, USA
    }

\begin{document}

\title{Fold-transversal surface code cultivation }%

\date{\today}

\author{Kaavya~Sahay}
\altaffiliation{These authors contributed equally to this work.}
\affiliation{\Yale}
\affiliation{\YQI}

\author{Pei-Kai~Tsai}
\altaffiliation{These authors contributed equally to this work.}
\affiliation{\Yale}
\affiliation{\YQI}

\author{Kathleen~(Katie)~Chang}
\altaffiliation{These authors contributed equally to this work.}
\affiliation{\Yale}
\affiliation{\YQI}

\author{Qile~Su}
\affiliation{\Yale}
\affiliation{\YQI}

\author{Thomas~B.~Smith}
\affiliation{\Yale}
\affiliation{\YQI}

\author{Shraddha~Singh}
\thanks{Present address: \IBM}
\affiliation{\Yale}
\affiliation{\YQI}

\author{Shruti~Puri}
\affiliation{\Yale}
\affiliation{\YQI}


\begin{abstract}
    Magic state cultivation is a state-of-the-art protocol to prepare ultra-high fidelity non-Clifford resource states for universal quantum computation.
    It offers a significant reduction in spacetime overhead compared to traditional magic state distillation techniques. 
    Cultivation protocols involve measuring a transversal logical Clifford operator on an initial small-distance code and then rapidly growing to a larger-distance code.
    In this work, we present a new cultivation scheme in which we measure the fold-transversal Hadamard of the unrotated surface code,
    and leverage unitary techniques to grow within the surface code family.
    Using both stabilizer and state vector simulations we find that this approach achieves the lowest known spacetime overhead for magic state cultivation. 
    Practical implementation of our protocol is best suited to architectures with non-local connectivity, showing the strength of architectures where such connectivity is readily available.
\end{abstract}


\maketitle

\section{Introduction}\label{sec:introduction}

Surface codes are a popular choice for quantum error-correction (QEC) due to their high thresholds and local, low-weight stabilizers~\cite{bravyi1998quantum,dennis2002topological,kitaev2003fault,fowler2012surface}.
A major bottleneck for scalable quantum computation with the surface code is the cost to implement fault-tolerant non-Clifford operations~\cite{fowler2013surface}. 
The most effective way to realize non-Clifford operations with these codes is by teleporting in high-fidelity magic states --- that is, an eigenstate of a Clifford operator~\cite{bravyi2005universal}. 
The logical error rate (LER) required for these magic states depends on the target quantum algorithm. 
For example, recent resource estimates for factoring large integers require input magic states with an infidelity of  roughly $10^{-7}$~\cite{gidney_factor_2025,zhou_resource_2025}.
The spacetime cost to reach such low LERs with surface codes can be debilitatingly large. 

Magic state cultivation (MSC) has recently emerged as an extremely resource-efficient method for preparing high-quality magic states~\cite{gidney2024magic,chen2025efficient,claes2025cultivatingtstatessurface,vaknin2025magic}. 
However, it requires a code that has a transversal implementation of a Clifford operator.
Since the rotated surface code does not have access to a transversal Clifford gate, MSC with the surface code needs to start in a different code that can later be morphed or grafted into a rotated surface code. 

MSC protocols proceed as follows.
Initially, the eigenstate of a Clifford operator is injected into a small-distance code.
This injection step is inherently noisy. 
Subsequently, the logical Clifford operator is measured transversally multiple times in order to `cultivate' a higher-fidelity magic state in the small code via postselection on non-trivial measurement outcomes. 
Finally, the small code is rapidly grown to a larger-distance rotated surface code for later use in a logical circuit. 
It is imperative that this final growth step does not significantly add to the LER of the protocol.
Therefore, further postselection is used in this step.

Note that the size of the initial code and number of logical Clifford measurements fixes the protocol's fault distance $\mathsf{f}$.
This is the minimum number of physical errors that may cause a logical error without introducing non-trivial syndromes. 
A protocol in which the Clifford operator is measured on a larger code has a higher achievable fault distance, and is thus capable of producing a higher-fidelity magic state.
However, this introduces increased spacetime overhead (STO) due to the use of additional qubits, more measurement rounds, and ultimately, more postselection.

\begin{figure*}
    \centering
    \includegraphics[page=1, width=\linewidth]{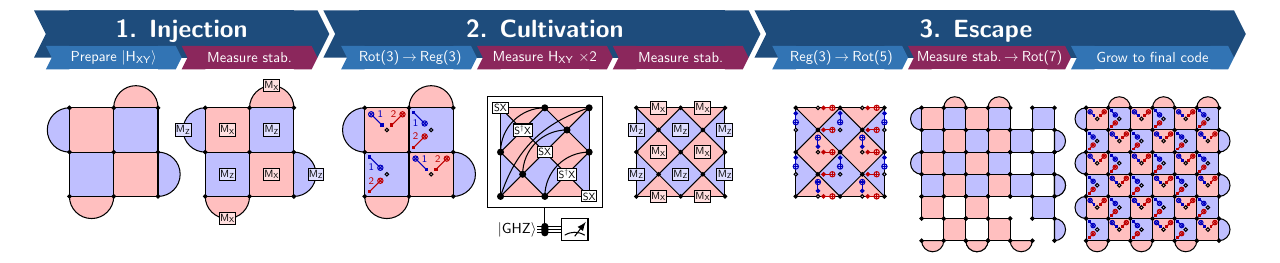}
    \caption{
        \textbf{Magic state cultivation} 
        with a fault distance $\fd3$.
        The procedure interleaves unitary steps (light blue labels) and measurements on which postselection occurs (maroon labels). 
        \textit{1. Injection:} 
        prepare an eigenstate of $H\XY = (X+Y)/\sqrt{2}$ on the distance-three rotated surface code $\mathsf{Rot}(3)$, and measure its stabilizers.
        \textit{2. Cultivation:} 
        transform to the distance-three regular surface code $\reg3$, and measure the fold-transversal $H\XY$ operator twice via a GHZ ancilla.
        Then, measure the stabilizers of $\reg3$.
        \textit{3. Escape:} 
        transform to $\rot5$ via unitary growth, then use stabilizer measurements to grow to $\rot7$. 
        Decode, and 
        post-select on the associated complementary gap. 
        If desired, unitarily grow to a larger final code.
        }
    \label{fig:summary}
\end{figure*}

\begin{figure*}
    \centering
    \begin{tikzpicture}
        \node at (0, 0) {
            \includegraphics[
                width = 0.5575\textwidth,
                trim = 0 10 0 10, 
                clip
                ]{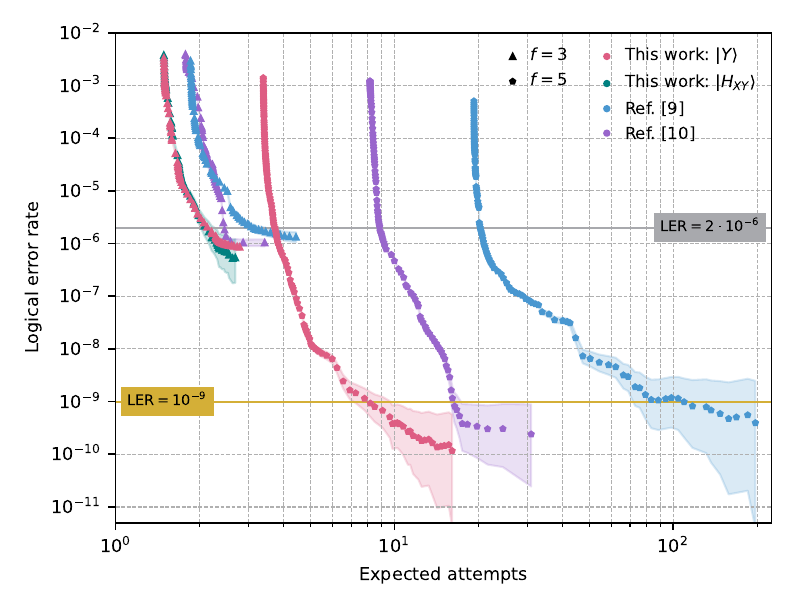}
                };
        \subfloat{\label{fig:LER}}%
        \node at (-3.35, 3.7) {\textbf{\textsf{(a)}}};
        \node at (8.9, 0.035) {
            \includegraphics[
                width = 0.4425\textwidth,
                trim = 0 10 0 10, 
                clip
                ]{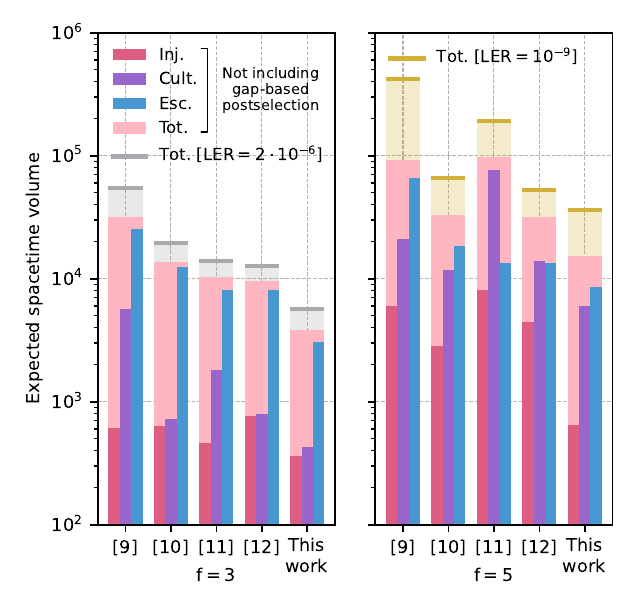}
                }; 
        \subfloat{\label{fig:STO}}%
        \node at (6.35, 3.7) {\textbf{\textsf{(b)}}};
    \end{tikzpicture}
    \caption{
        \textbf{Cultivation under uniform depolarizing noise.} 
        \textbf{(a)} 
        Curves show the expected number of cultivation attempts of fault distances $\fd3$ (triangles) and $\fd5$ (pentagons) to prepare a state with a particular logical error rate.
       The $\fd3$ protocol is described in~\cref{sec:protocol}, and its extension to $\fd5$ in~\cref{app:fd5}.
        Our scheme for $\ket{Y}$ state cultivation (dark pink) requires a lower number of expected attempts for any target LER compared to Refs.~\cite{gidney2024magic,chen2025efficient} (blue and purple, respectively).
        The exact simulation of $\ket{H\XY}$ magic state cultivation for $\fd3$ (green) is in good agreement with $\ket{Y}$ state results. 
        \textbf{(b)
        Expected} spacetime volume (qubits $\times$ gate count $\times$ expected attempts) of $\ket{Y}$ state cultivation for fault distances $\fd3$ and $\fd5$, compared to Refs.~\cite{gidney2024magic,chen2025efficient,claes2025cultivatingtstatessurface,vaknin2025magic}.
        For a fair comparison, the $\fd3$ protocols are normalized to the same final code distance $d\fin=13$, and the $\fd5$ protocols to $d\fin=15$.
        The expected volume in the absence of gap-based postselection ({light pink}) is shown, along with the component expected volumes for injection (dark pink), cultivation (purple) and escape (blue).
        Additional gap-based postselection is required to cultivate states with logical error rates of $2\times 10^{-6}$ for $\fd3$ (silver) and $10^{-9}$ for $\fd5$ (gold).
        Our scheme has a significantly lower overhead, owing to the efficient cultivation and escape steps that use nonlocal operations.
        }
    \label{fig:e2e}
\end{figure*}

In this work, we present a new method for cultivating a high-fidelity $H\XY = (X+Y)/\sqrt{2}$ state in the rotated surface code.
Our protocol is summarized in~\cref{fig:summary}. 
We observe that the unrotated surface code has a fold-transversal logical $H\XY$ gate~\cite{kubica2015unfolding,moussa2016transversal}. 
We leverage this property for cultivation in the unrotated surface code, along with a method for unitarily transforming between rotated and unrotated surface code variants~\cite{tsai2025unitary}.  
Our choice of initial code, magic state, and simplified growth technique allow us to achieve lower logical error rates and significantly lower spacetime overheads compared to previous schemes. 

Importantly, our scheme is most naturally implemented with non-local gates.
These are native to quantum computing platforms such as neutral atoms~\cite{bluvstein2024logical} and trapped ions~\cite{moses2023racetrack}. 
More recently, high-fidelity non-local gates have also been demonstrated in superconducting circuits~\cite{wang2025demonstration,qldpc2025cross}. 
Our work demonstrates the dramatic reduction in requirements for scalable quantum computing with surface codes when non-local gates are available. 

The performance of our magic state cultivation scheme is summarized by~\cref{fig:LER}.
This is quantified by the expected attempts required to obtain a magic state of a particular LER.
Here, our results are compared to existing
cultivation protocols for which the equivalent data was available~\cite{gidney2024magic,chen2025efficient}.
We find that our procedure requires a lower number of expected attempts for any target LER.
The overhead relative to all current protocols~\cite{gidney2024magic,chen2025efficient,vaknin2025magic,claes2025cultivatingtstatessurface} is shown in~\cref{fig:STO}.
This is measured by the expected spacetime volume.
We calculate the `baseline' expected volume not accounting for the additional postselection after the escape stage --- as well as the expected volume required to obtain a magic state with an error rate of $2 \times 10^{-6}$ and $10^{-9}$.
The use of non-local operations allows us to reduce the expected volume
of injection, cultivation, and escape.
As such, we obtain a much lower STO compared to existing work.

The remainder of this paper is structured as follows.
In~\cref{sec:protocol} we describe our MSC protocol in detail.
In~\cref{sec:results} we benchmark our protocol against previous work by simulating cultivation of the logical $Y$ eigenstate, $\ket{Y} = S\ket{+}$.
In addition, we use a combination of
state vector and stabilizer simulations to accurately benchmark cultivation for the {logical magic state}, $\ket{H\XY} = T\ket{+}$.
We also investigate the performance of our protocol under physically-motivated noise models.
We conclude in~\cref{sec:conclusion}.

\section{Protocol}\label{sec:protocol}

Our MSC protocol prepares a high-fidelity eigenstate of the logical operator $H\XY = (X+Y)/\sqrt{2}$ in a rotated surface code.
We label magic states by the operator for which they are a $+1$ eigenstate. 
Here, the magic state is $\ket{H\XY} =(\ket{0}+e^{i\pi/4}\ket{1})/\sqrt{2}$, which is used to implement the logical $T =e^{-i\pi Z/8}$ gate via teleportation.
\Cref{fig:summary} illustrates our protocol for cultivating the $\ket{H\XY}$ state with a fault distance $\fd3$. 
The protocol has three main stages: injection, cultivation and escape.

\textit{Injection---} 
In this stage, {we prepare a} logical magic state in a small-distance rotated surface code, $\rot3$.
While there are several ways to achieve this, 
here we use a unitary encoding circuit, followed by a round of stabilizer measurements.
The encoding circuit, which uses one physical non-Clifford $T$ gate and non-local Clifford operations, is provided in~\cref{app:msc-circuits}. 
This injection stage has a fault distance $\fd1$.

\textit{Cultivation---} 
In this stage, another unitary circuit is used to transform $\rot3$ into an unrotated surface code, $\reg3$. 
This circuit is simply the first two steps of a stabilizer measurement round on $\rot3$~\cite{mcewen2023relaxing}.
Note that this technique requires the introduction of additional qubits to act as the ancillae for $\reg3$.
How this is achieved depends on underlying physical hardware.
It is most naturally realized in architectures with reconfigurable qubits, such as neutral atom arrays~\cite{bluvstein2024logical}.

After this transformation, the fold-transversal logical $H\XY$ operator is measured twice using a three-qubit ancilla GHZ state. 
This operation uses some three-qubit gates, as shown in~\cref{app:msc-circuits}. 
Unlike previous proposals~\cite{gidney2024magic, vaknin2025magic, chen2025efficient}, our logical Clifford measurement procedure is designed to minimize the wait time for code qubits while operations are being carried out only on the GHZ ancillae.

Following this logical Clifford measurement, we perform a round of stabilizer measurements at $\reg3$. 
Post-selecting on both the stabilizer measurements and the decoded GHZ measurements increases the fault distance of the protocol to $\fd3$. 
At this point, we have produced a high-fidelity magic state on a small-distance code. 

\textit{Escape---}
In this stage, the code is grown to a larger distance in order to protect the magic state from future errors. 
Here, we use a three-step escape strategy. 
First, $\reg3$ is grown to $\rot5$ via a local unitary circuit~\cite{tsai2025unitary}.
Next, $\rot5$ is grown to $\rot7$ using a conventional stabilizer measurement approach~\cite{li2015magic}.
Finally, $\rot7$ is unitarily grown to a final larger-distance code.  
The stabilizer measurements obtained from the growth to $\rot7$ are used for additional postselection.
Shots are discarded depending on the likelihood of a logical error as determined by a complementary-gap-based decoder~\cite{gidney2025yoked,bombin2024fault,hutter2014efficient,meister2024efficient}.

This three-stage escape allows us to do most of our growth unitarily, while still extracting the stabilizer information necessary for gap-based postselection.
The reliance on primarily unitary circuits is crucial for maintaining a low spacetime overhead compared to using possibly slower and lower-fidelity stabilizer measurements on the larger-distance final code.
 
We provide further details about the protocol in the appendices.
In~\cref{app:msc-circuits} we provide circuits for key steps, and in~\cref{app:alts} we discuss alternative choices for each stage.
These choices include: the magic state, injection strategy, GHZ ancilla size, pre-escape stabilizer measurement code, and escape strategy. 
In~\cref{app:fd5} we outline the additional steps required for protocols with a higher fault distance.
These are a straightforward extension of the $\fd3$ protocol that we have described above.
Note that we must measure the logical $H\XY$ operator additional times to achieve a higher fault distance.
Other modifications to the $\fd5$ scheme include a modified logical Clifford measurement circuit, and the use of a local, unitary circuit~\cite{claes2025lowerdepthlocalencodingcircuits} for escape.

\begin{table*}
    \begin{tabular}{| c | c | c | c | c | c | c |}
        \hline
        \textbf{Scheme} & \textbf{Initial} & \textbf{Clifford check} & \textbf{Clifford check} & \textbf{Discard rate of} & \textbf{Performance constraints}
        \\
        & \textbf{code} &  \textbf{code}  & \textbf{ancilla state} & \textbf{\boldsymbol{$\ket{Y}$} with LER \boldsymbol{$=10^{-9}$}} &
        \\
        \hline Ref.~\cite{gidney2024magic} & Color(3) & Color  &  $\ket{+}^{\otimes 19}$  & $99\%$& Planar connectivity, grafting 
        \\ Ref.~\cite{chen2025efficient} & $\mathbb{RP}^2-3$ & $\mathrm{SRP}$ &  $\ket{+}^{\otimes 25}$  & $93\%$ & Inefficient code morphing  
        \\ Ref.~\cite{claes2025cultivatingtstatessurface} & $\rot3$ & $S-\mathsf{Reg}$ & $\ket{+}^{\otimes 9}$ & $99\%$ & Restricted connectivity, $\leq 2$-qubit gates
        \\ Ref.~\cite{vaknin2025magic} & $\rot3$ & $\mathsf{Reg}$  &  $\ket{+}^{\otimes 21}$ & $96\%$& High-overhead logical check 
        \\ This work & $\rot3$ & $\mathsf{Reg}$ &  GHZ(5)  & $90\%$ & 
        \\
        \hline
    \end{tabular}
    \caption{
        \textbf{Comparison of cultivation schemes} for $\ket{H\XY}$ with fault distance $\fd5$.
        \textit{Initial code:} the code chosen for state injection. 
        \textit{Clifford check code}: the code in which logical Clifford checks are performed. 
        \textit{Clifford check ancilla}: the largest ancillary system used to perform the logical Clifford check, where $\mathrm{GHZ}(n)=(\ket{0}^{\otimes n}+\ket{1}^{\otimes n})/\sqrt{2}$.
        \textit{Discard rate:} the {expected} discard rate required to {cultivate a $\ket{Y}$ state with} LER~$= 10^{-9}$, where discard rate ${1-}1/(\mathrm{expected~attempts})$. 
        \textit{Performance constraints:} {aspects of each cultivation scheme that} limit their performance relative to ours.  
        {Note that all schemes end in a rotated surface code.}
        }
    \label{tbl:prot-comp}
\end{table*}

\subsection{Comparison with other schemes}

A summary of the differences between our protocol and previous schemes is provided in~\cref{tbl:prot-comp}. 
Here we highlight the constraints of previous schemes that limit their relative
LER and STO.

Ref.~\cite{gidney2024magic} uses a technique called grafting to convert between a color code to the rotated surface code in the escape stage~\footnote{Ref.~\cite{gidney2024magic} first escapes into a `matchable' code that aims to be as similar to $\mathsf{Rot}$ as possible.}. 
This introduces additional postselection overhead that we bypass by remaining in the $\mathsf{Rot}$-$\mathsf{Reg}$ surface code family.
Refs.~\cite{chen2025efficient,claes2025cultivatingtstatessurface} use morphing circuits to convert, respectively, between $\mathbb{RP}^2$ and $\mathsf{Reg}$ codes and their self-dual variants.
This enables a strictly transversal logical Clifford measurement using two-qubit gates. 
In contrast, we relax the requirement of morphing to a self-dual code and allow for the use of three-qubit gates. 
Our code conversion within the surface code family is {comparatively} more efficient. 
Ref.~\cite{claes2025cultivatingtstatessurface} only allows operations that can be efficiently realized on a neutral-atom platform under reconfigurability constraints. While we consider this as a design principle, we leave exact optimization to future work. 

{\Cref{fig:STO} shows that we are able to obtain the lowest expected spacetime volume for both all stages: injection, cultivation, and escape.}
We attribute most of our improvements to the use of a more efficient check circuit, 
and our three-stage escape strategy.
These improvements are discussed further throughout~\cref{app:alts}.

\section{Results}\label{sec:results}

In this section, we quantify the performance of our protocol numerically. 
In~\cref{sec:comparison}, we use 
stabilizer simulations of logical $\ket{Y}$ state preparation to benchmark against previous works. 
Using the $\ket{Y}$ state as a proxy for the $\ket{H\XY}$ state avoids simulating non-Clifford circuits. 
However, it is yet unclear if this provides robust estimates of the LER of magic state cultivation. 
Motivated by this, in~\cref{sec:error-estimation} we introduce a \textit{handoff} approach, where outputs from a state vector simulator are passed to a stabilizer simulator. 
This allows us to study exact $\ket{H\XY}$ state cultivation for the first time.
Finally, in~\cref{sec:nonlocal}, we examine the performance of our scheme in an experimentally-motivated setting.

\subsection{\texorpdfstring{$\ket{Y}$}{Y} state cultivation}\label{sec:comparison}

To adapt our scheme for the logical $\ket{Y}$ state, we make two changes to the protocol described in~\cref{sec:protocol}. 
First, injection is modified to use an $S$ gate instead of a $T$ gate~\cite{gidney2019efficient,chen2025efficient}.
Second, we measure the logical $Y$ operator in place of measuring the logical $H\XY$ operator. 
We provide the relevant circuits in~\cref{app:msc-circuits}.
In this section, we compare our scheme against other cultivation schemes
in terms of LER vs. expected attempts,
and expected spacetime volume~\cite{gidney2024magic,chen2025efficient}. 
Circuits for all schemes are simulated under a uniform depolarizing noise model (SD6) with physical error rate $p = 10^{-3}$.

\Cref{fig:LER} shows the expected attempts to achieve a particular LER.
For each curve, the leftmost data point reflects the performance without any complementary-gap-based postselection during escape. 
As one progresses right, increasing amounts of postselection via the complementary gap drives the LER lower, until the error floor is reached by fully post-selecting on any nontrivial syndrome. 
For any given LER, our protocol consistently requires lower expected attempts than existing proposals.
Additionally, our scheme is the only one so far for which $\fd3$ cultivation has a floor below $10^{-6}$. 
This brings us slightly closer to the $7 \times 10^{-7}$ LER required for 2048 bit integer factorization~\cite{zhou_resource_2025}.
For our $\fd5$ cultivation, {10 expected attempts are required to} reach an LER of $10^{-9}$, fewer than the 15--90 attempts required by previous proposals. 
 
\Cref{fig:STO} highlights the advantage of our approach in terms of overhead.
We calculate the expected spacetime volume required for target LERs of $2\times 10^{-6}$ {and $10^{-9}$}, for protocols with fault distances $\fd3$ and $\fd5$, respectively.
The spacetime volume is defined as the qubit–gate product (the number of active qubits times the number of gate steps). 
This metric naturally accounts for differences in depth and qubit count across protocols, and also reflects the fact that error detection can halt a protocol mid-way rather than incurring the cost of executing the full circuit. 
Using this metric provides a more accurate comparison than the expected number of attempts.
Our protocol demonstrates significant improvement in spacetime volume compared to Refs.~\cite{gidney2024magic,chen2025efficient}. 
We refer the reader to~\cref{app:sims} for further details on the calculation of expected spacetime volume. 

\subsection{\texorpdfstring{$\ket{H\XY}$}{H} state cultivation}\label{sec:error-estimation}
 
Here, we describe the state vector to stabilizer handoff simulation that is used to accurately estimate the LER for $\ket{H\XY}$ magic state cultivation.
In~\cref{app:handoff} we describe our handoff simulation strategy in detail.
We provide a high-level summary below.

First, we use a state vector simulator from injection up to the completion of the logical Clifford measurements.
This part of the circuit involves non-Clifford physical gates. 
but only involves a limited number of qubits. 
After the logical measurements, the logical qubit state vector and the code's stabilizer eigenvalues are extracted. 
This logical qubit state vector is stored offline and the stabilizer eigenvalues are passed onto the stabilizer simulator. 

The stabilizer simulator then simulates the rest of the protocol.
The decoder is only given stabilizer measurement information post{-}handoff and predicts Pauli errors that have occurred.
If the predicted errors differ from the actual errors by logical Pauli operators, then the corresponding logical Pauli operator is applied to the stored pre-handoff 
logical qubit state vector. 
We compare this final noisy state to the ideal magic state to determine if the protocol has succeeded. 

\Cref{fig:LER} contains the results for $\ket{H\XY}$ cultivation with $\fd3$. 
Interestingly, the LERs for $\ket{H\XY}$ cultivation are virtually indistinguishable from those for $\ket{Y}$ state cultivation. 
Ref.~\cite{gidney2024magic} observed a difference between the LERs of $\ket{Y}$ and $\ket{H\XY}$ cultivation when simulating a truncated protocol, where the escape stage is excluded.
We also observe a difference for truncated simulations, but this effect does not carry over to our full simulations that include escape.
We discuss $\ket{Y}$ vs. $\ket{H\XY}$ simulations further in~\cref{app:handoff}. 

The state vector simulations are only performed up to the logical checks in the cultivation stage, which involve a small number of qubits.
This renders handoff tractable for the $\fd3$ protocol.  
However, our handoff strategy is too expensive to simulate $\fd5$, which uses roughly 47 qubits during cultivation.
Nevertheless, our demonstration for $\fd3$ establishes it as a useful tool for estimating the performance of protocols where the system is limited to a small number of qubits during non-Clifford operations in a circuit.

\subsection{Tailoring cultivation to experiments}\label{sec:nonlocal}

Finally, we examine the performance of our protocol under
a noise model inspired by non-local architectures such as neutral atoms. 
Broadly speaking, in quantum systems with non-local operations, single-qubit gates are less noisy than two-qubit gates, and 
local gates are less noisy than non-local gates. 
Moreover, for neutral atoms, idle errors are an insignificant portion of the error budget. 
For simplicity, we simulate the preparation of $\ket{Y}$ as, based on the results of the previous section, we expect 
its trends to extend to $\ket{H\XY}$ 
cultivation.

We consider two physically{-}motivated noise models, denoted PM1 and PM5.
In these noise models, there are no idle errors, single-qubit gates have
uniform depolarizing noise at a rate $p/10$, and local multi-qubit gates 
have uniform depolarizing noise at a rate $p$.
The two models differ in the noise strength assigned to non-local multi-qubit gates. 
In PM1 these gates experience uniform depolarizing noise at a rate $p$ (the same as local multi-qubit gates), and in PM5 they have a higher noise rate $5p$.
We also use a more natural gate set for neutral atom qubits, and decompose CX gates into CZ gates and local single-qubit rotations. 

Note that cultivation requires the state to be in the $+1$ eigenspace of all stabilizers. 
If using measurement-based injection, this necessitates fast measurement-conditioned feedback, which may not be experimentally desirable. 
Here, we use optimized unitary state preparation followed by a subset of stabilizer measurements. 
Importantly, fast feedback based on the stabilizer measurements is not required in this injection strategy, which is described in more detail in~\cref{app:alts}. 

\begin{figure}[t]
    \centering
    \includegraphics[
        width = \linewidth,
        trim = 0 10 0 10, 
        clip
        ]{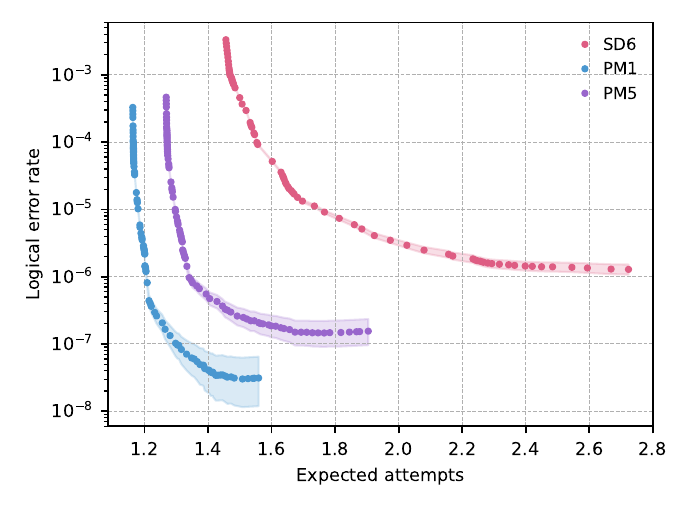}
    \caption{ 
        \textbf{Physically-motivated noise.}
        Logical error rate of $\ket{Y}$ state cultivation under three different noise models.
        A standard depolarizing noise model (SD6) is compared to two physically-motivated noise models (PM1 and PM5).
        In PM1 the noise strength of single-qubit gates is ten times lower than that of multi-qubit gates.
        The same is true for PM5, but the noise strength of non-local multi-qubit gates is five times that of local multi-qubit gates.
        PM1 and PM5 have no idle errors.
        }
    \label{fig:metricomps}
\end{figure}

\Cref{fig:metricomps} shows the performance of our 
protocol under these noise models with $p=10^{-3}$, compared to the standard depolarizing noise model (SD6) used in previous sections.
We observe that PM noise models leads to reduced LERs compared to SD6 even when non-local gates are heavily penalized. 
This indicates that the protocol is largely limited by single-qubit gates and idling errors.
Furthermore, our protocol requires less expected attempts for a given LER compared to Ref.~\cite{vaknin2025magic}.

\section{Discussion and Conclusion}\label{sec:conclusion}

In this work, we have presented a fold-transversal cultivation scheme for the surface code.
We have benchmarked its performance by evaluating $\ket{Y}$ state cultivation, and for the first time have studied exact $\ket{H\XY}$ magic state cultivation. 
Our protocol achieves the lowest known spacetime overhead for MSC.
This improved performance is due to efficient unitary circuits that allows us to stay entirely within the surface code family throughout, and utilize the native fold-transversal $H\XY$ operator.
Both of these tasks require non-local operations.
There are some natural extensions of our work. 

The first arises from the reduced sampling cost of our cultivation protocol.
As our scheme reaches lower LERs with far fewer expected attempts, it is conceivable to simulate fault distances higher than $\fd5$. 
This may enable us to determine the necessary spacetime overhead to produce a magic state with much lower LER, perhaps low enough that further distillation is not even required. 

It is also important to study the performance of MSC under realistic noise models.
While we only consider Pauli errors in our work, real hardware experiences a variety of other error mechanisms.  
It would be informative to analyze performance of our scheme in the presence of erasure errors~\cite{vaknin2025magic,jacoby2025magicstateinjectionerasure}, which is relevant to neutral atom architectures~\cite{wu2022erasure,ma2023high,scholl2023erasure,zhang2025leveragingerasureerrorslogical}. 
Due to the ability to post-select on erasure, we anticipate further improved fidelity in this setting.
Several platforms also experience leakage or coherent errors.
However, these are more challenging to simulate.

Throughout this work we used a non-local unitary circuit to grow the surface code, and convert between rotated and unrotated variants.
This contributes to our lower spacetime overheads. 
However, non-local unitary growth is not compatible with all qubit platforms. 
It is worthwhile to consider low-overhead growth using alternative schemes such as the local unitary circuits of Ref.~\cite{higgott2021optimal,claes2025lowerdepthlocalencodingcircuits}. 
These circuits have an inherent tradeoff between gate depth and idling time. 
If an overhead reduction is possible via their use, then our protocol may be better applied to fixed-qubit architectures with limited non-local connectivity. 

Beyond these immediate extensions of our work, an avenue of future research is to extend magic state cultivation to other quantum low-density parity check (LDPC) codes hosting fold-transversal gates~\cite{breuckmann2024fold} and families with morphing circuits~\cite{shaw2025lowering}.

We expect our work to inspire new surface code protocols that leverage experimental advances in non-local operations for overhead reduction. 
At the very least, these advances call for a more fair comparison between the surface code and other quantum LDPC codes that have non-local stabilizers. 
Now equipped with a full computational suite of accurate decoders, efficient Clifford operations, and low-overhead magic state generation in non-local architectures, surface codes remain a leading candidate for practical quantum computation.

\section*{Acknowledgments}

We thank Ludwig Schmid for modifying the unitary encoding optimizers of Ref.~\cite{mqt2024}, which led to the identification of the optimized unitary injection scheme. 
We would also like to thank Craig Gidney, Jeff D. Thompson, and Jahan Claes for helpful discussions, and the authors of Refs.~\cite{gidney2023cleaner,gidney2024magic, gidney2025yoked}, for releasing their code, which has been used extensively in this work.
This work was supported under  ARO (grant no. W911NF-23-1-0051),  DARPA MeasQuIT (grant no. HR00112490363), NSF QLCI (grant no. OMA2120757), and DOE C$^2$QA (contract number DE-SC0012704).

\textit{Related work---} 
The authors were made aware of two related publications in the same arXiv posting: an updated version of Ref.~\cite{vaknin2025magic} with an investigation of $\ket{H}$ state cultivation on the surface code, and  surface code cultivation using two-qubit gates from J. Claes~\cite{claes2025cultivatingtstatessurface}.

\textit{Data availability---} sample code and data for this work are provided at Ref.~\cite{gitrepo}. 

\bibliography{refs}

@article{breuckmann2024fold,
    title={{Fold-Transversal Clifford Gates for Quantum Codes}},
    volume={8},
    url = {https://doi.org/10.22331/q-2024-06-13-1372},
    journal={Quantum},
    publisher={Verein zur Forderung des Open Access Publizierens in den Quantenwissenschaften},
    author={Breuckmann, Nikolas P. and Burton, Simon},
    year={2024},
    month=jun, 
    pages={1372},
}

@article{moussa2016transversal,
    title = {{Transversal Clifford gates on folded surface codes}},
    author = {Moussa, Jonathan E.},
    journal = {Phys. Rev. A},
    volume = {94},
    issue = {4},
    pages = {042316},
    numpages = {14},
    year = {2016},
    month = {Oct},
    publisher = {American Physical Society},
    url = {https://link.aps.org/doi/10.1103/PhysRevA.94.042316}
}

@misc{gidney2023cleaner,
    title={{Cleaner magic states with hook injection}}, 
    author={Craig Gidney},
    year={2023},
    eprint={2302.12292},
    archivePrefix={arXiv},
    primaryClass={quant-ph},
}

@misc{gidney2024magic,
    title={{Magic state cultivation: growing T states as cheap as CNOT gates}}, 
    author={Craig Gidney and Noah Shutty and Cody Jones},
    year={2024},
    eprint={2409.17595},
    archivePrefix={arXiv},
    primaryClass={quant-ph},
}

@article{chamberland2020very,
    title={{Very low overhead fault-tolerant magic state preparation using redundant ancilla encoding and flag qubits}},
    volume={6},
    ISSN={2056-6387},
    url={http://dx.doi.org/10.1038/s41534-020-00319-5},
    number={1},
    journal={npj Quantum Information},
    publisher={Springer Science and Business Media LLC},
    author={Chamberland, Christopher and Noh, Kyungjoo},
    year={2020},
    month=oct 
}

@article{li2015magic,
    title={{A magic state’s fidelity can be superior to the operations that created it}},
    volume={17},
    ISSN={1367-2630},
    url={http://dx.doi.org/10.1088/1367-2630/17/2/023037},
    number={2},
    journal={New Journal of Physics},
    publisher={IOP Publishing},
    author={Li, Ying},
    year={2015},
    month=feb, 
    pages={023037},
}

@article{higgott2021optimal,
    title={{Optimal local unitary encoding circuits for the surface code}},
    volume={5},
    ISSN={2521-327X},
    journal={Quantum},
    publisher={Verein zur Forderung des Open Access Publizierens in den Quantenwissenschaften},
    author={Higgott, Oscar and Wilson, Matthew and Hefford, James and Dborin, James and Hanif, Farhan and Burton, Simon and Browne, Dan E.},
    year={2021},
    month=aug, 
    pages={517},
    url = {https://doi.org/10.22331/q-2021-08-05-517},
}

@article{fowler2012surface,
    title = {{Surface codes: Towards practical large-scale quantum computation}},
    author = {Fowler, Austin G. and Mariantoni, Matteo and Martinis, John M. and Cleland, Andrew N.},
    journal = {Phys. Rev. A},
    volume = {86},
    issue = {3},
    pages = {032324},
    numpages = {48},
    year = {2012},
    month = {Sep},
    publisher = {American Physical Society},
    url = {https://link.aps.org/doi/10.1103/PhysRevA.86.032324}
}

@misc{wang2025demonstration,
      title={{Demonstration of low-overhead quantum error correction codes}}, 
      author={Ke Wang and Zhide Lu and Chuanyu Zhang and Gongyu Liu and Jiachen Chen and Yanzhe Wang and Yaozu Wu and Shibo Xu and Xuhao Zhu and Feitong Jin and Yu Gao and Ziqi Tan and Zhengyi Cui and Ning Wang and Yiren Zou and Aosai Zhang and Tingting Li and Fanhao Shen and Jiarun Zhong and Zehang Bao and Zitian Zhu and Yihang Han and Yiyang He and others},
      year={2025},
      eprint={2505.09684},
      archivePrefix={arXiv},
      primaryClass={quant-ph},
}

@misc{tsai2025unitary,
      title={{A Unitary Encoder for Surface Codes}}, 
      author={Pei-Kai Tsai and Shruti Puri},
      year={2025},
      eprint={2506.04084},
      archivePrefix={arXiv},
      primaryClass={quant-ph},
}

@misc{vaknin2025magic,
    title={{Efficient Magic State Cultivation on the Surface Code}}, 
    author={Yotam Vaknin and Shoham Jacoby and Arne Grimsmo and Alex Retzker},
    year={2025},
    eprint={2502.01743},
    archivePrefix={arXiv},
    primaryClass={quant-ph},
}

@misc{chen2025efficient,
    title={{Efficient Magic State Cultivation on {$\mathbb{RP}^2$}}}, 
    author={Zi-Han Chen and Ming-Cheng Chen and Chao-Yang Lu and Jian-Wei Pan},
    year={2025},
    eprint={2503.18657},
    archivePrefix={arXiv},
    primaryClass={quant-ph},
}

@article{lao2022magic,
    title={{Magic state injection on the rotated surface code}},
    author={Lingling Lao and Ben Criger},
    journal={Proceedings of the 19th ACM International Conference on Computing Frontiers},
    year={2022},
    url={https://doi.org/10.1145/3528416.3530237}
}

@article{bluvstein2024logical,
    title={{Logical quantum processor based on reconfigurable atom arrays}},
    author={Bluvstein, Dolev and Evered, Simon J and Geim, Alexandra A and Li, Sophie H and Zhou, Hengyun and Manovitz, Tom and Ebadi, Sepehr and Cain, Madelyn and Kalinowski, Marcin and Hangleiter, Dominik and others},
    journal={Nature},
    volume={626},
    number={7997},
    pages={58--65},
    year={2024},
    publisher={Nature Publishing Group UK London},
    url={https://doi.org/10.1038/s41586-023-06927-3}
}

@misc{lima_clifford_2025,
    title={{Clifford and Non-Clifford Splitting in Quantum Circuits: Applications and ZX-Calculus Detection Procedure}}, 
    author={Fernando Lima and Arcesio Castañeda Medina},
    year={2025},
    eprint={2504.16004},
    archivePrefix={arXiv},
    primaryClass={quant-ph},
}

@article{kubica2015unfolding,
   title={Unfolding the color code},
   volume={17},
   ISSN={1367-2630},
   url={http://dx.doi.org/10.1088/1367-2630/17/8/083026},
   DOI={10.1088/1367-2630/17/8/083026},
   number={8},
   journal={New Journal of Physics},
   publisher={IOP Publishing},
   author={Kubica, Aleksander and Yoshida, Beni and Pastawski, Fernando},
   year={2015},
   month=aug, pages={083026} }

@article{daguerre_code_2025,
    title = {{Code Switching Revisited: Low-overhead Magic State Preparation Using Color Codes}},
    shorttitle = {Code Switching Revisited},
    author = {Daguerre, Lucas and Kim, Isaac H.},
    year = {2025},
    month = apr,
    journal = {Physical Review Research},
    volume = {7},
    number = {2},
    pages = {023080},
    publisher = {American Physical Society},
    doi = {10.1103/PhysRevResearch.7.023080},
    urldate = {2025-05-20},
}

@misc{zhou_resource_2025,
    title={{Resource Analysis of Low-Overhead Transversal Architectures for Reconfigurable Atom Arrays}}, 
    author={Hengyun Zhou and Casey Duckering and Chen Zhao and Dolev Bluvstein and Madelyn Cain and Aleksander Kubica and Sheng-Tao Wang and Mikhail D. Lukin},
    year={2025},
    eprint={2505.15907},
    archivePrefix={arXiv},
}

@misc{gidney_factor_2025,
    title={{How to factor 2048 bit RSA integers with less than a million noisy qubits}}, 
    author={Craig Gidney},
    year={2025},
    eprint={2505.15917},
    archivePrefix={arXiv},
}

@article{mcewen2023relaxing,
    title={{Relaxing hardware requirements for surface code circuits using time-dynamics}},
    author={McEwen, Matt and Bacon, Dave and Gidney, Craig},
    journal={Quantum},
    volume={7},
    pages={1172},
    year={2023},
    publisher={Verein zur F{\"o}rderung des Open Access Publizierens in den Quantenwissenschaften},
    url = {https://doi.org/10.22331/q-2023-11-07-1172},
}

@inproceedings{mqt2024,
    title={{The MQT Handbook: A Summary of Design Automation Tools and Software for Quantum Computing}},
    url={http://dx.doi.org/10.1109/QSW62656.2024.00013},
    booktitle={2024 IEEE International Conference on Quantum Software (QSW)},
    publisher={IEEE},
    author={Wille, Robert and Berent, Lucas and Forster, Tobias and Kunasaikaran, Jagatheesan and Mato, Kevin and Peham, Tom and Quetschlich, Nils and Rovara, Damian and Sander, Aaron and Schmid, Ludwig and Schönberger, Daniel and Stade, Yannick and Burgholzer, Lukas},
    year={2024},
    month=jul, 
    pages={1–8} 
}

@article{gidney2019efficient,
    title={{Efficient magic state factories with a catalyzed {CCZ} to 2{T} transformation}},
    author={Gidney, Craig and Fowler, Austin G},
    journal={Quantum},
    volume={3},
    pages={135},
    year={2019},
    publisher={Verein zur F{\"o}rderung des Open Access Publizierens in den Quantenwissenschaften},
    url = {https://doi.org/10.22331/q-2019-04-30-135},
}

@misc{jacoby2025magicstateinjectionerasure,
    title={{Magic State Injection with Erasure Qubits}}, 
    author={Shoham Jacoby and Yotam Vaknin and Alex Retzker and Arne L. Grimsmo},
    year={2025},
    eprint={2504.02935},
    archivePrefix={arXiv},
    primaryClass={quant-ph},
}

@article{singh2022high,
    title={{High-fidelity magic-state preparation with a biased-noise architecture}},
    author={Singh, Shraddha and Darmawan, Andrew S and Brown, Benjamin J and Puri, Shruti},
    journal={Physical Review A},
    volume={105},
    number={5},
    pages={052410},
    year={2022},
    publisher={APS},
    doi = {10.1103/PhysRevA.105.052410},
}

@article{bombin2024fault,
    title={{Fault-tolerant postselection for low-overhead magic state preparation}},
    author={Bomb{\'\i}n, H{\'e}ctor and Pant, Mihir and Roberts, Sam and Seetharam, Karthik I},
    journal={PRX Quantum},
    volume={5},
    number={1},
    pages={010302},
    year={2024},
    publisher={APS},
    url = {https://link.aps.org/doi/10.1103/PRXQuantum.5.010302}
}

@article{hutter2014efficient,
    title={{Efficient Markov chain Monte Carlo algorithm for the surface code}},
    author={Hutter, Adrian and Wootton, James R and Loss, Daniel},
    journal={Physical Review A},
    volume={89},
    number={2},
    pages={022326},
    year={2014},
    publisher={APS},
    url = {https://link.aps.org/doi/10.1103/PhysRevA.89.022326}
}

@misc{meister2024efficient,
    title={{Efficient soft-output decoders for the surface code}}, 
    author={Nadine Meister and Christopher A. Pattison and John Preskill},
    year={2024},
    eprint={2405.07433},
    archivePrefix={arXiv},
    primaryClass={quant-ph},
}

@misc{zhang2025leveragingerasureerrorslogical,
    title={{Leveraging erasure errors in logical qubits with metastable $^{171}$Yb atoms}}, 
    author={Bichen Zhang and Genyue Liu and Guillaume Bornet and Sebastian P. Horvath and Pai Peng and Shuo Ma and Shilin Huang and Shruti Puri and Jeff D. Thompson},
    year={2025},
    eprint={2506.13724},
    archivePrefix={arXiv},
    primaryClass={quant-ph},
}

@article{wu2022erasure,
    title={{Erasure conversion for fault-tolerant quantum computing in alkaline earth Rydberg atom arrays}},
    author={Wu, Yue and Kolkowitz, Shimon and Puri, Shruti and Thompson, Jeff D},
    journal={Nature communications},
    volume={13},
    number={1},
    pages={4657},
    year={2022},
    publisher={Nature Publishing Group UK London},
    doi={10.1038/s41467-022-32094-6}
}

@article{scholl2023erasure,
    title={{Erasure conversion in a high-fidelity Rydberg quantum simulator}},
    author={Scholl, Pascal and Shaw, Adam L and Tsai, Richard Bing-Shiun and Finkelstein, Ran and Choi, Joonhee and Endres, Manuel},
    journal={Nature},
    volume={622},
    number={7982},
    pages={273--278},
    year={2023},
    publisher={Nature Publishing Group UK London},
    doi={10.1038/s41586-023-06516-4}
}

@article{ma2023high,
    title={{High-fidelity gates and mid-circuit erasure conversion in an atomic qubit}},
    author={Ma, Shuo and Liu, Genyue and Peng, Pai and Zhang, Bichen and Jandura, Sven and Claes, Jahan and Burgers, Alex P and Pupillo, Guido and Puri, Shruti and Thompson, Jeff D},
    journal={Nature},
    volume={622},
    number={7982},
    pages={279--284},
    year={2023},
    publisher={Nature Publishing Group UK London},
    doi={10.1038/s41586-023-06438-1}
}

@article{kitaev2003fault,
    title={{Fault-tolerant quantum computation by anyons}},
    author={Kitaev, A Yu},
    journal={Annals of physics},
    volume={303},
    number={1},
    pages={2--30},
    year={2003},
    publisher={Elsevier},
    url = {https://www.sciencedirect.com/science/article/pii/S0003491602000180},
}

@article{dennis2002topological,
    title={{Topological quantum memory}},
    author={Dennis, Eric and Kitaev, Alexei and Landahl, Andrew and Preskill, John},
    journal={Journal of Mathematical Physics},
    volume={43},
    number={9},
    pages={4452--4505},
    year={2002},
    publisher={American Institute of Physics},
    url = {https://doi.org/10.1063/1.1499754}, 
}

@misc{bravyi1998quantum,
      title={{Quantum codes on a lattice with boundary}}, 
      author={S. B. Bravyi and A. Yu. Kitaev},
      year={1998},
      eprint={quant-ph/9811052},
      archivePrefix={arXiv},
}

@article{fowler2013surface,
    title={{Surface code implementation of block code state distillation}},
    author={Fowler, Austin G and Devitt, Simon J and Jones, Cody},
    journal={Scientific reports},
    volume={3},
    number={1},
    pages={1939},
    year={2013},
    publisher={Nature Publishing Group UK London},
    url = {https://doi.org/10.1038/srep01939}
}

@article{bravyi2005universal,
    title={{Universal quantum computation with ideal Clifford gates and noisy ancillas}},
    author={Bravyi, Sergey and Kitaev, Alexei},
    journal={Physical Review A},
    volume={71},
    number={2},
    pages={022316},
    year={2005},
    publisher={APS},
    url = {https://link.aps.org/doi/10.1103/PhysRevA.71.022316},
}

@article{shaw2025lowering,
    title = {{Lowering Connectivity Requirements for Bivariate Bicycle Codes Using Morphing Circuits}},
    author = {Shaw, Mackenzie H. and Terhal, Barbara M.},
    journal = {Phys. Rev. Lett.},
    volume = {134},
    issue = {9},
    pages = {090602},
    numpages = {5},
    year = {2025},
    month = {Mar},
    publisher = {American Physical Society},
    doi = {10.1103/PhysRevLett.134.090602},
}

@article{gidney2025yoked,
    title={{Yoked surface codes}},
    author={Gidney, Craig and Newman, Michael and Brooks, Peter and Jones, Cody},
    journal={Nature Communications},
    url={https://doi.org/10.1038/s41467-025-59714-1},
    volume={16},
    number={1},
    pages={4498},
    year={2025},
    publisher={Nature Publishing Group UK London}
}

@article{moses2023racetrack,
    title = {{A Race-Track Trapped-Ion Quantum Processor}},
    author = {Moses, S. A. and Baldwin, C. H. and Allman, M. S. and Ancona, R. and Ascarrunz, L. and Barnes, C. and Bartolotta, J. and Bjork, B. and Blanchard, P. and Bohn, M. and Bohnet, J. G. and Brown, N. C. and Burdick, N. Q. and Burton, W. C. and Campbell, S. L. and Campora, J. P. and others},
    journal = {Phys. Rev. X},
    volume = {13},
    issue = {4},
    pages = {041052},
    numpages = {25},
    year = {2023},
    month = {Dec},
    publisher = {American Physical Society},
    doi = {10.1103/PhysRevX.13.041052},
}

@misc{gitrepo,
    author  = {Kaavya Sahay and Pei-Kai Tsai and Katie Chang},
    title   = {{Fold transversal surface code cultivation}},
    url     = {https://github.com/kaavyas99/MSC_foldedH},
    year    = {2025},
    note    = {Accessed: 04 September 2025} 
}

@inproceedings{qldpc2025cross,
    title={{A modular quantum computer based on bivariate bicycle codes}},
    url={https://yale.hosted.panopto.com/Panopto/Pages/Viewer.aspx?id=28b64187-d5f5-4eae-bbd3-b33000fc6aba},
    booktitle={{7th International Conference on Quantum Error Correction}},    
    author={Cross, Andrew},
    year={2025},
    month=aug
}

@article{sparse25higgott,
  doi = {10.22331/q-2025-01-20-1600},
  url = {https://doi.org/10.22331/q-2025-01-20-1600},
  title = {{Sparse {B}lossom: correcting a million errors per core second with minimum-weight matching}},
  author = {Higgott, Oscar and Gidney, Craig},
  journal = {{Quantum}},
  issn = {2521-327X},
  publisher = {{Verein zur F{\"{o}}rderung des Open Access Publizierens in den Quantenwissenschaften}},
  volume = {9},
  pages = {1600},
  month = jan,
  year = {2025}
}

@misc{gidney_sinter,
    title = {{Sinter}: fast {QEC} sampling},
    author = {Gidney, Craig},
    howpublished = {\url{https://pypi.org/project/sinter/}},
    note = {Accessed: 2025-09-05},
    year = {2025},
}

@misc{claes2025cultivatingtstatessurface,
    title={{Cultivating T states on the surface code with only two-qubit gates}}, 
    author={Jahan Claes},
    year={2025},
    eprint={2509.05232},
    archivePrefix={arXiv},
    primaryClass={quant-ph},
}

@misc{claes2025lowerdepthlocalencodingcircuits,
      title={Lower-depth local encoding circuits for the surface code}, 
      author={Jahan Claes},
      year={2025},
      eprint={2509.09779},
      archivePrefix={arXiv},
      primaryClass={quant-ph},
      url={https://arxiv.org/abs/2509.09779}, 
}

\appendix
\crefalias{section}{appendix}

\setcounter{equation}{0}
\setcounter{figure}{0}
\setcounter{table}{0}

\renewcommand{\thefigure}{A\arabic{figure}}
\renewcommand{\thetable}{A\arabic{table}}

\section{Circuits}\label{app:msc-circuits}

\begin{figure*}
    \newpage
    \includegraphics[width=0.95\linewidth]{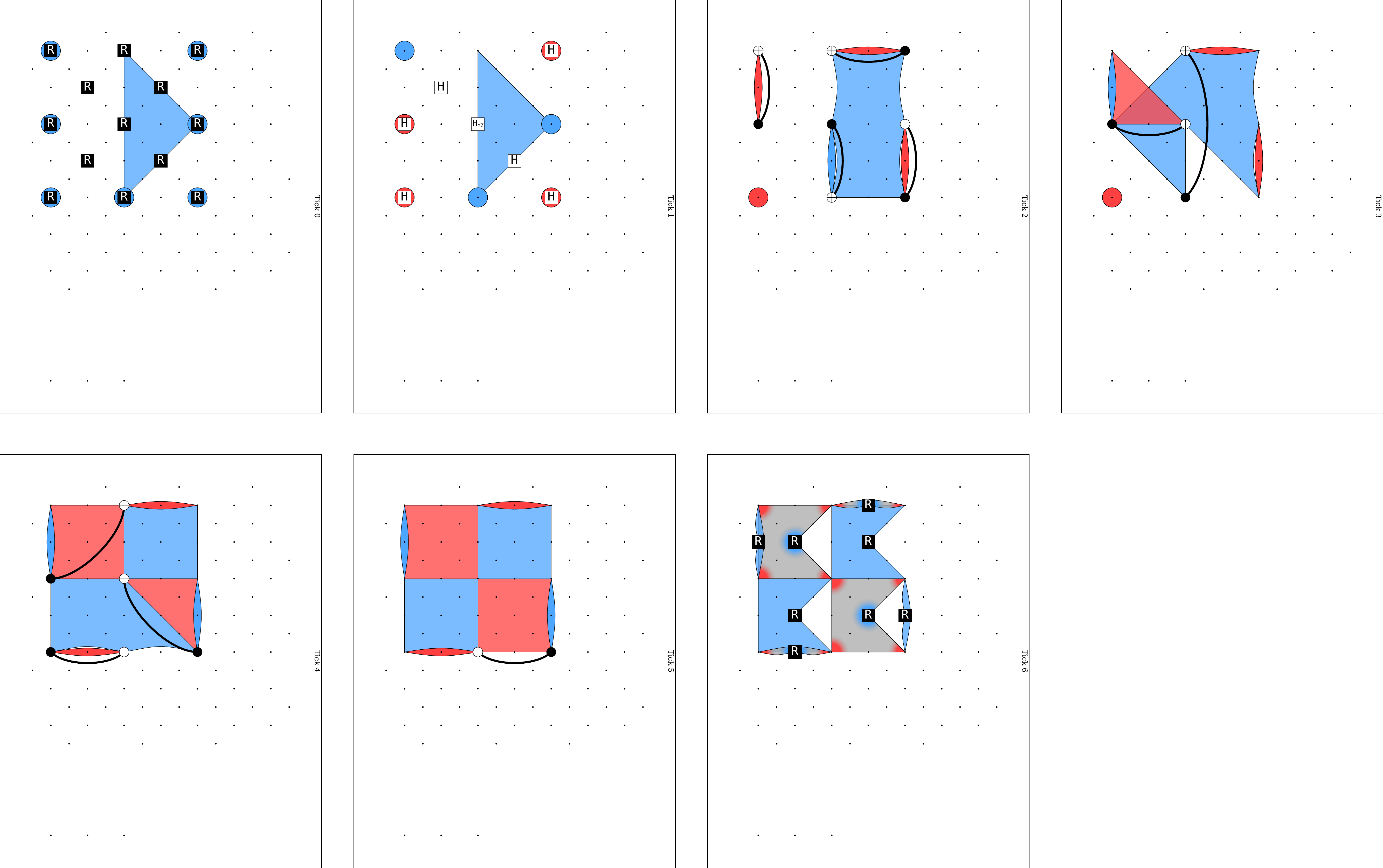}
    \caption{
        \textbf{Unitary encoder for \boldsymbol{$\rot3$}.}
        {The rotation on the central qubit in the second step produces a physical $\ket{Y}$ state. 
        In subsequent steps this is used to unitarily prepare the logical $\ket{Y}$ state.}
        A subsequent round of stabilizer measurements (not pictured) is used to detect weight-1 and 2 errors.
        }
    \label{fig:unit-inj-s}
\end{figure*}

\begin{figure*}
    \includegraphics[width=0.95\linewidth]{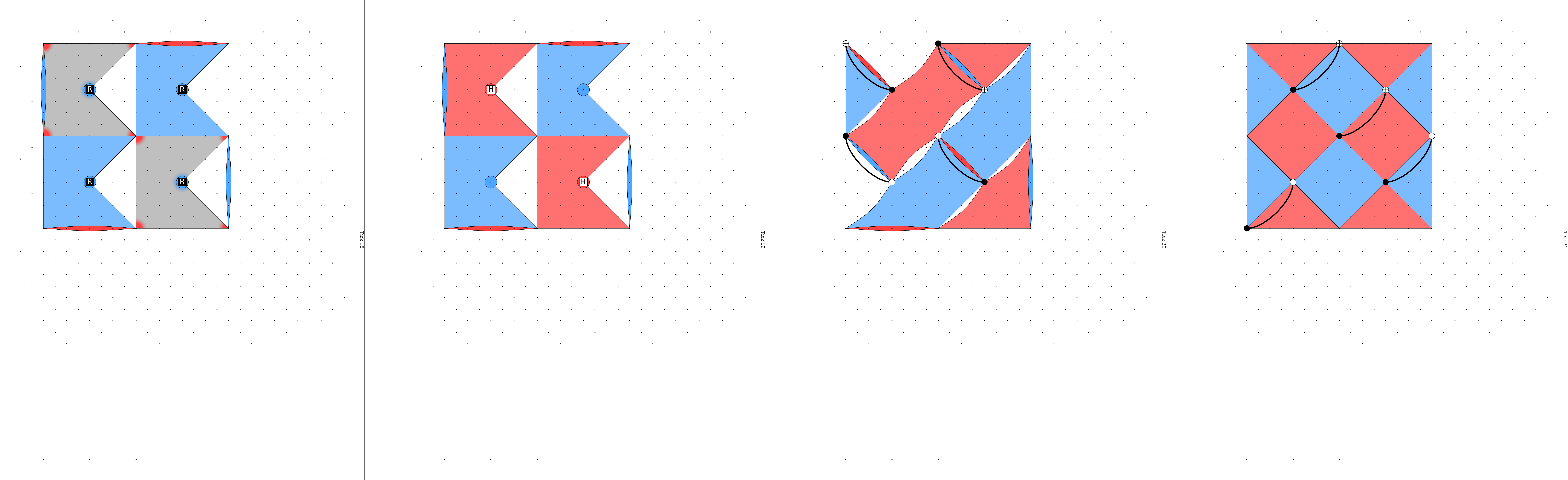}
    \caption{
        \textbf{Growth from \boldsymbol{$\rot3$} to \boldsymbol{$\reg3$}.}
        This is simply the first two steps of the stabilizer measurement circuit for $\rot3$.
        The ancilla qubits of $\rot3$ become data qubits of $\reg3$.
        Additional qubits are required for the $\reg3$ code for stabilizer measurements.
        Depending on the hardware, these ancilla may be present beforehand (static architectures with non-local connectivity), or freshly initialized qubits may be introduced to the lattice (reconfigurable architectures).
        }
    \label{fig:rotreg-f3}
\end{figure*}

\begin{figure*}
    \includegraphics[page=2,width=0.5\linewidth]{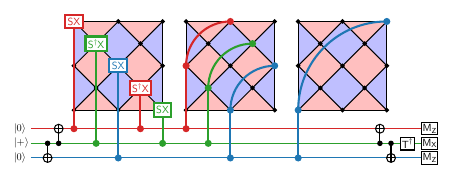}
    \caption{
        \textbf{Logical \boldsymbol{$H\XY$} check for \boldsymbol{$\reg3$}.}
        The ancilla GHZ(3) state is supported on the bottom wires. 
        The check is decomposed into controlled-$SX$ gates involving the diagonal qubits, and CCZ gates involving pairs of off-diagonal qubits. 
        }
    \label{fig:hxy-check-f3}
\end{figure*}

\begin{figure*}
    \includegraphics[width=0.95\linewidth]{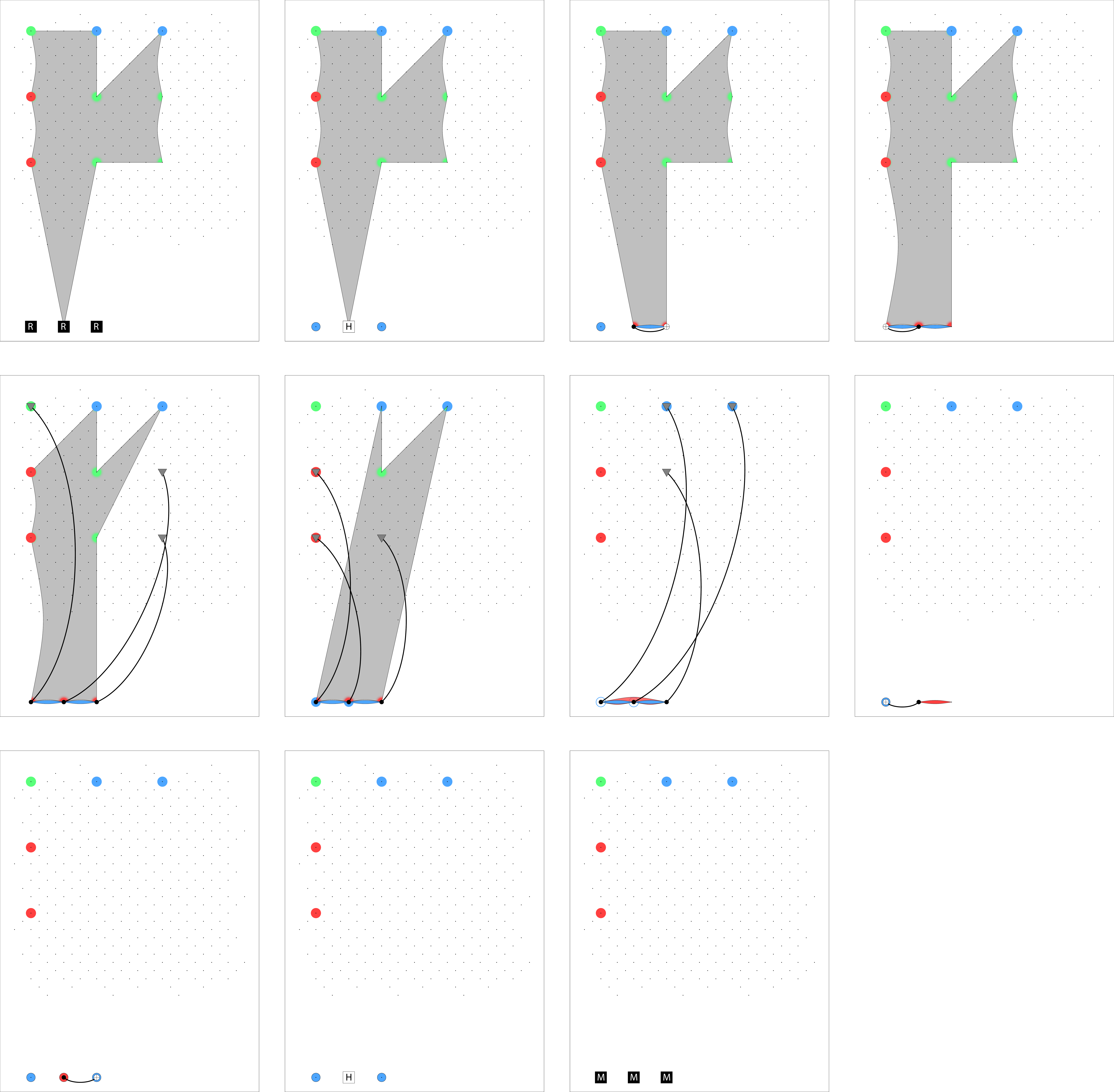}
    \caption{
        \textbf{Logical \boldsymbol{$Y$} check for \boldsymbol{$\reg3$}.}
        The logical observable is highlighted. 
        Note that since the logical $Y$ operator
        is transversal on $\reg3$ (compared to the fold-transversal 
        logical $H\XY$ operator) we only need two-qubit gates.
        }
    \label{fig:y-check-f3}
\end{figure*}

\begin{figure*}
    \includegraphics[width=0.95\linewidth]{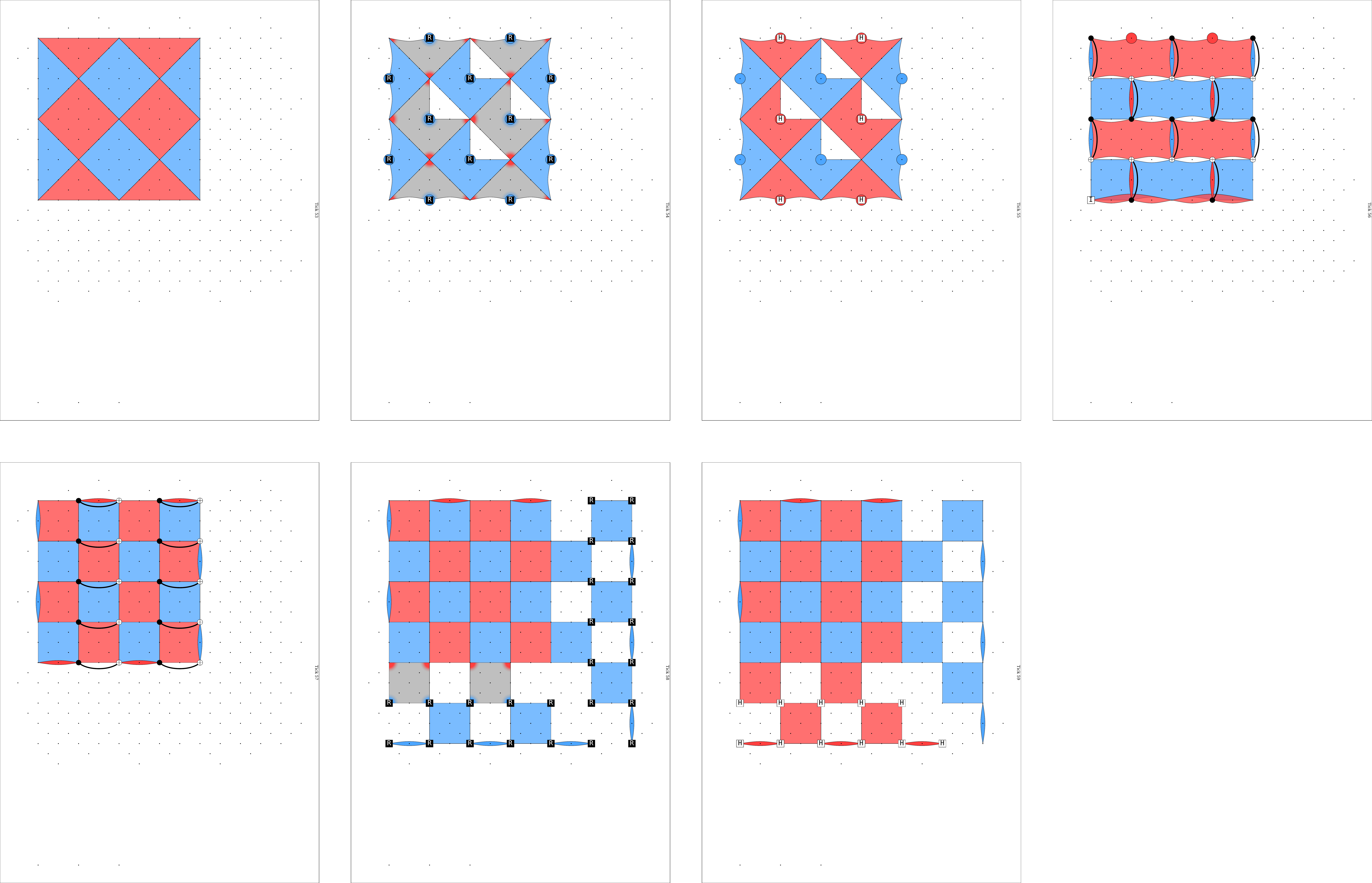}
    \caption{ 
        \textbf{Unitary growth from \boldsymbol{$\reg3$} to \boldsymbol{$\rot5$}.}
        This is the first stage of our three-stage escape sequence. 
        The first two steps of stabilizer measurement circuit for $\reg3$ are used to grow and rotate aka \textit{growtate} the code to $\rot5$.
        As in~\cref{fig:rotreg-f3}, the ancilla qubits of $\reg3$ become data qubits of $\rot5$, and additional ancilla qubits need to be introduced.
        Furthermore, two patches of freshly initialized qubits, used for the second stage of escape, are shown.
        This second stage involves subsequent rounds of stabilizer measurements (not pictured) that induce growth from $\rot5$ to $\rot7$.
        }
    \label{fig:growth1-f3}
\end{figure*}

\begin{figure*}
    \includegraphics[width=0.7\linewidth]{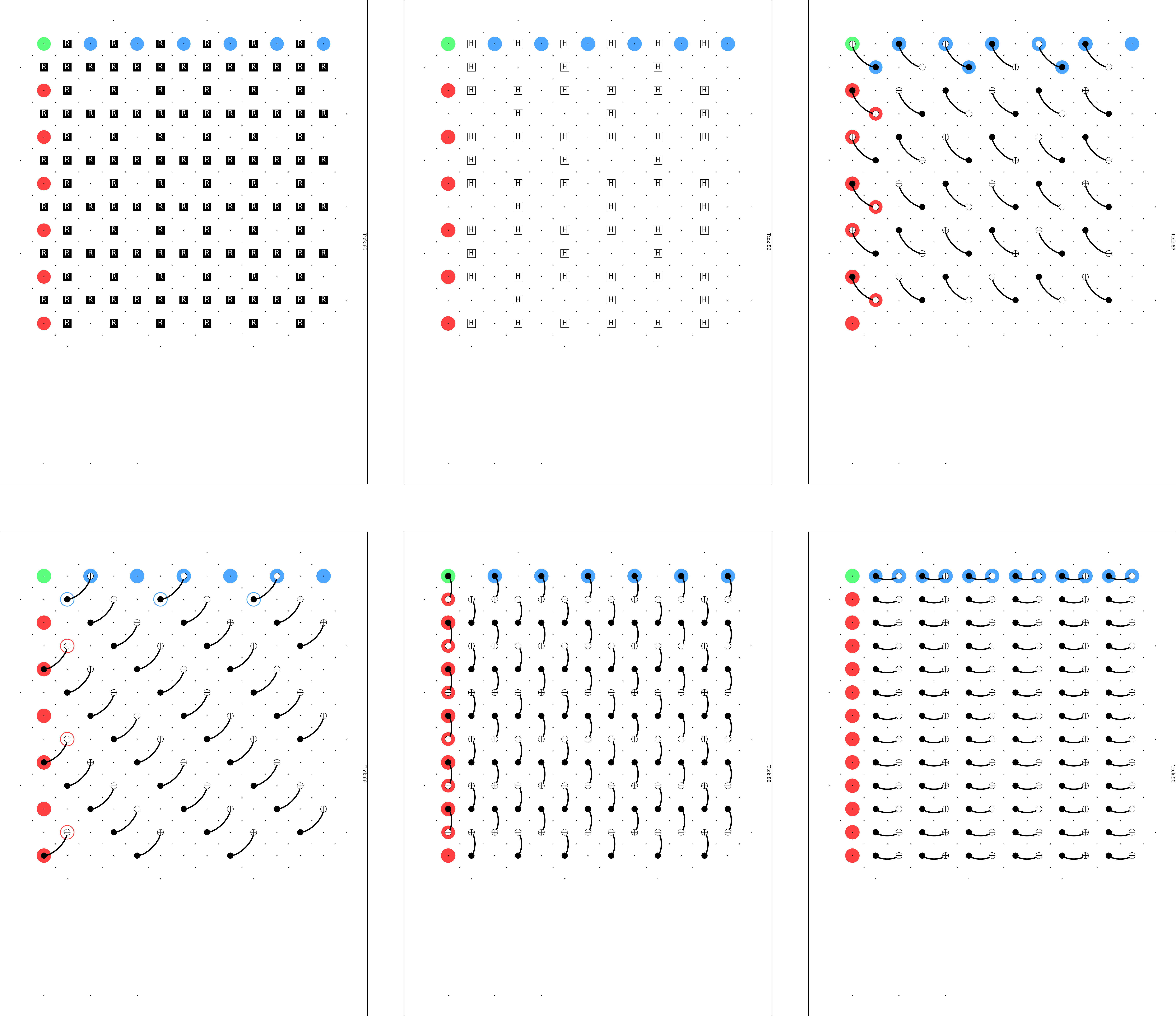}
    \caption{
        \textbf{Unitary growth from \boldsymbol{$\rot7$} to \boldsymbol{$\rot{13}$}.}
        Detectors are omitted for visual clarity.
        This is the final stage of our three-stage escape sequence.
        The logical $Y$ observable is highlighted, showing the increase in its support.
        }
    \label{fig:growth2-f3}
\end{figure*}

In this section, we provide circuit diagrams for key stages of our protocol.
The circuits used to 
simulate fault distance $\fd3$ $\ket{Y}$ state cultivation 
(\cref{fig:e2e}) are shown in~\cref{fig:unit-inj-s,fig:rotreg-f3,fig:y-check-f3,fig:growth1-f3,fig:growth2-f3} (generated in \texttt{Stim}).
Red regions in these figures represent $X$ detectors (which flip in the presence of $Z$ errors), and blue regions represent $Z$ detectors (which are affected by $X$ errors). 
{The $H\XY$ check circuit for $\reg3$ is shown in~\cref{fig:hxy-check-f3}.
This circuit is used to simulate $\ket{H\XY}$ state cultivation {(\cref{fig:e2e})}.}

\begin{figure*}
    \includegraphics[page=3,width=\linewidth]{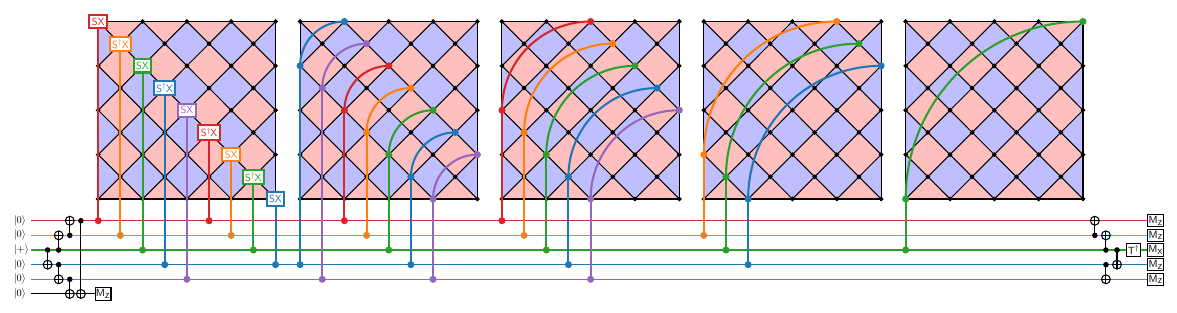}
    \caption{
        \textbf{Logical \boldsymbol{$Y$} check for \boldsymbol{$\reg 5$}.}
        The ancilla GHZ(5) state is supported on the bottom wires, 
        with an additional qubit used as a flag for fault-tolerant preparation. 
        This circuit ensures that four or less errors on the GHZ state do not propagate to an undetectable logical error on the code.
        }
    \label{fig:hxy-check-f5}
\end{figure*}

In~\cref{app:fd5} we describe how these circuits are modified for state cultivation with a higher fault distance $\fd5$.
A notable modification is the design of 
the logical check circuit for $\reg{5}$.
In~\cref{fig:hxy-check-f5} we show the $H\XY$ check circuit, and in~\cref{fig:f5-Y-check} we show the equivalent $Y$ check circuit.

Finally, an alternative unitary encoding circuit
is shown in~\cref{fig:optunit-f3}{.} 
This is used to simulate $\ket{Y}$ state cultivation under physically-motivated constraints (\cref{fig:metricomps}).

\begin{figure*}
    \centering
    \includegraphics[width=0.95\linewidth]{f5_checkcy.pdf}
    \caption{
        \textbf{Logical \boldsymbol{$Y$} check for \boldsymbol{$\reg 5$}.}
        }
    \label{fig:f5-Y-check}
\end{figure*}

\begin{figure*}
    \includegraphics[width=0.95\linewidth]{appb_optunit.pdf}
    \caption{
        \textbf{Alternative unitary encoder for \boldsymbol{$\rot3$}.}
        This circuit only requires the subsequent measurement of three code stabilizers (detectors shown) to maintain fault distance.
        It is therefore particularly suited to architectures where measurements are more costly than gates in terms of speed, fidelity or both.
        }
    \label{fig:optunit-f3}
\end{figure*}

\clearpage

\section{Potential modifications} \label{app:alts}

In~\cref{sec:protocol}, we described a
cultivation protocol that achieved both low logical error rates and reduced spacetime overhead.
This protocol is highly modular, where any given stage can be modified to use different codes or subroutines. 
Here, we describe several 
alternatives, showing the resultant performance tradeoffs.
This discussion is structured by going through the procedure for cultivation from beginning to end and considering points where modifications may be made.
In this section and the remainder of the appendix, logical operators $\bar O$ are denoted with a bar, to differentiate them from physical operators $O$.

\subsection{Logical state} \label{app:HXY-encode}

In this section, we discuss the logical non-Clifford state used for magic state cultivation.
We restrict ourselves to considering protocols where the logical state is checked at $\reg3$. 
In this setting, checking a Hadamard-like operator presents a natural choice due to its fold-transversality on $\reg3$. 
Here, we briefly digress to first prove the correctness of our implementation of the logical $\bar{H}\XY$ on $\reg3$, and then show that  $\bar{H}\XY$ requires less physical gates compared to 
$\bar{H}_\mathrm{XZ}$. 

In unrotated surface codes, the logical $\bar{S}$ can be implemented fold-transversally~\cite{moussa2016transversal,breuckmann2024fold}. 
For a $d\times d$ unrotated surface code, let $D_i$ be the $i$-th qubit along the main diagonal where $0\leq i \leq 2d-2$, $\Delta$ be the set of qubits below the main diagonal (not included), and $\tau$ be the map that reflects a qubit across the main diagonal. 
For a qubit $q\in\Delta$, we call $(q, \tau(q))$ a pair of \textit{mirrored qubits}. 
$\bar{S}$ can be implemented as
\begin{equation}
    \bar{S} 
    = 
    \left( S_{D_0} S\dg_{D_1} \cdots S\dg_{D_{2d-3}} S_{D_{2d-2}} \right) 
    \prod_{q \in \Delta} CZ_{q, \tau(q)}
    ,
\end{equation}
where $S$ and $S\dg$ gates alternately act on qubits on the main diagonal, and CZs are applied to each pair of mirrored qubits.

As $\bar{X}$ can be applied fold-transversally by applying $X$ on all qubits on the main diagonal, 
$\bar{H}\XY = e^{-i\pi/4} \bar{S} \bar{X}$ is {also} fold-transversal. 
Specifically, $\bar{H}\XY$ can be written as
\begin{equation}
    \begin{aligned} 
        \label{eq: HXY details}
        \bar{H}\XY
        &= 
        e^{-i\pi/4} \bar{S} \bar{X} 
        \\
        &= 
        e^{-i\pi/4} 
        \prod_{i\text{ even}}(SX)_{D_i} 
        \prod_{j\text{ odd}}(S\dg X)_{D_j} 
        \prod_{q \in \Delta} CZ_{q, \tau(q)} 
        \\
        &= 
        \prod_{i\text{ even}}(TXT\dg)_{D_i} 
        \prod_{j\text{ odd}}(TYT\dg)_{D_j}  
        \prod_{q \in \Delta} CZ_{q, \tau(q)}
        \\
        &= 
        \prod_{i\text{ even}}(G_X Z G_X\dg)_{D_i}  
        \prod_{j\text{ odd}}(G_Y Z G_Y\dg)_{D_j} 
        \prod_{q \in \Delta} CZ_{q, \tau(q)}
    \end{aligned}
\end{equation}
where $G_X=TH$, $G_Y=TSH$ are single-qubit rotations, and we have used the gate identity $TXT\dg = e^{-i\pi/4}SX$. 

To implement a controlled-$\bar{H}\XY$ gate for logical measurements, we need a CZ gate (conjugated by $G_X$ or $G_Y$) for each qubit on the main diagonal, and a CCZ gate for each pair of mirrored qubit. 
In total, there are $(2d-1)$ CZ gates and $(d-1)^2$ CCZ gates.

The unrotated surface code supports a fold-transversal
$\bar{H}_\mathrm{XZ} = (\bar{X} + \bar{Z})/\sqrt 2$, whose +1 eigenstate $\ket{\bar{H}_\mathrm{XZ}}$ is also a magic state that is Clifford-equivalent to the conventional $\ket T$ state. 
It is implemented as 
\begin{equation}
    \bar{H}_\mathrm{XZ}
    =
    \left(\prod_{q \in \Delta} \mathrm{SWAP}_{q, \tau(q)} \right) H^{\otimes n}
    ,
\end{equation}
where $H^{\otimes n}$ denotes physical Hadamard gates on all qubits, followed by SWAP gates on each pair of mirrored qubits. 
To measure $\bar{H}_\mathrm{XZ}$, its controlled version involves CH (controlled-Hadamard) gates acting on all data qubits, and CSWAP (controlled-SWAP) gates on each pair of mirrored qubit. 
Each CH can be written as $R_Y(\pi/4)\cdot CZ\cdot R_Y(-\pi/4)$, and each CSWAP gate can be decomposed into a single CCZ gate and two CZ gates. 
In total, there are $(d^2+3(d-1)^2)$ CZ gates and $(d-1)^2$ CCZ gates. 
The number of three-qubit gates is the same as in $\bar{H}\XY$, with an increase in the number of two-qubit gates, as summarized in~\cref{table:GateCounts}. 
Therefore we choose to prepare the $\ket{\bar{H}\XY}$ state in this work.

\begin{table}
\centering
    \begin{tabular}{|c|c|c|c|}
        \hline
        $\boldsymbol{d}$ & $\boldsymbol{n_\mathrm{2Q}}$ \textbf{for} $\boldsymbol{\bar{H}\XY}$ & $\boldsymbol{n_\mathrm{2Q}}$ \textbf{for} $\boldsymbol{\bar{H}_\mathrm{XZ}}$ & $\boldsymbol{n_\mathrm{3Q}}$ \\
        \hline
        $3$ & 5 & 21 & 4 \\
        \hline
        $5$ & 9 & 73 & 16 \\
        \hline
        $d$ & $2d-1$ & $4d^2-6d+3$ & $(d-1)^2$\\
        \hline
    \end{tabular}
    \caption{
        \textbf{Gate count for controlled-\boldsymbol{$\bar H$} checks.}
        We show the number of two-qubit gates ($n_\mathrm{2Q}$) and three-qubit gates ($n_\mathrm{3Q}$) for different code distances $d$.
        }
    \label{table:GateCounts}
\end{table}

We implement the controlled-$\bar{H}\XY$ using a combination of controlled-$SX$ and CCZ gates.
\Cref{fig:hxy-check-f3} shows the $\bar H\XY$ check circuit for $\fd3$.
Since the decomposition $\bar H\XY = e^{-i\pi/4} \bar S \bar X$ introduces an extra phase factor, an additional $T\dg$ gate must be applied to the ancilla qubit. 
This gate is placed immediately before the ancilla $X$-basis measurement.
\Cref{fig:hxy-check-f5} shows the $\bar H\XY$ check for $\mathsf f=5$.
In practice, one may prefer the decomposition using CZ, CCZ, and single-qubit gates as in~\cref{eq: HXY details}. 
In this form, no $T\dg$ gate on the ancilla is required; instead, suitable single-qubit rotations are applied before and after the CZ gates.

\subsection{Injection scheme}\label{app:injection}

In this section, we discuss different ways to inject the $\ket{\bar{H}\XY}$ state.
Note that we restrict ourselves to injection into $\rot3$ and not $\reg3$ due to its lower spacetime overhead. 
We consider the following:

\begin{figure*}
    \centering
    \begin{tikzpicture}
        \node at (0, 0) {
            \includegraphics[
                height = 0.4\textwidth,
                trim = 0 10 0 10, 
                clip
                ]{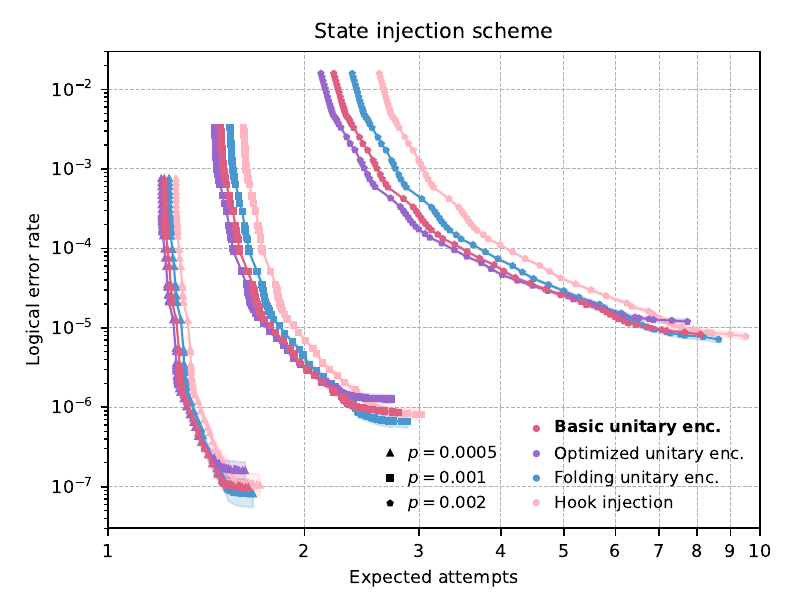}
                };
        \subfloat{\label{fig:injvary}}%
        \node at (-3.4, 3.45) {\textbf{\textsf{(a)}}};
        \node at (0, -7.5) {
            \includegraphics[
                height = 0.4\textwidth,
                trim = 0 10 0 10, 
                clip
                ]{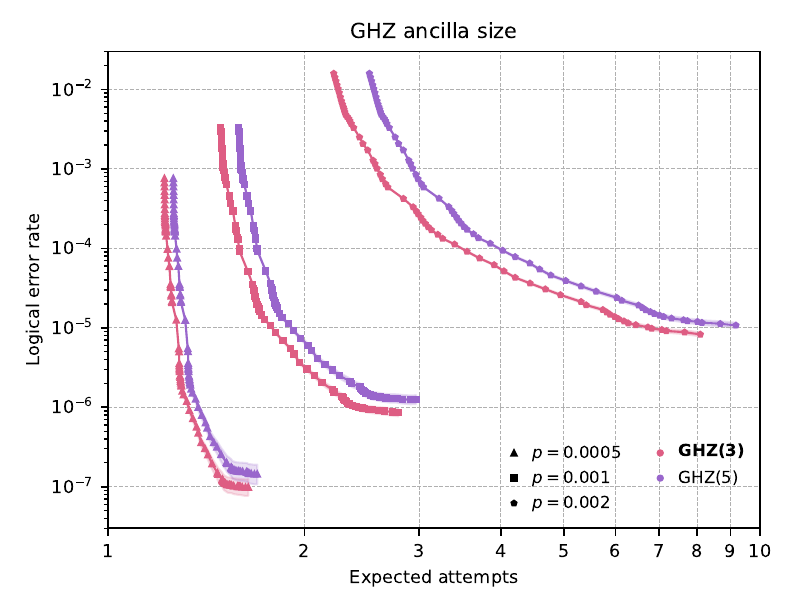}
                }; 
        \subfloat{\label{fig:ghzvary}}%
        \node at (-3.4, 3.45-7.5) {\textbf{\textsf{(b)}}};
        \node at (8.9, 0) {
            \includegraphics[
                height = 0.4\textwidth,
                trim = 0 10 0 10, 
                clip
                ]{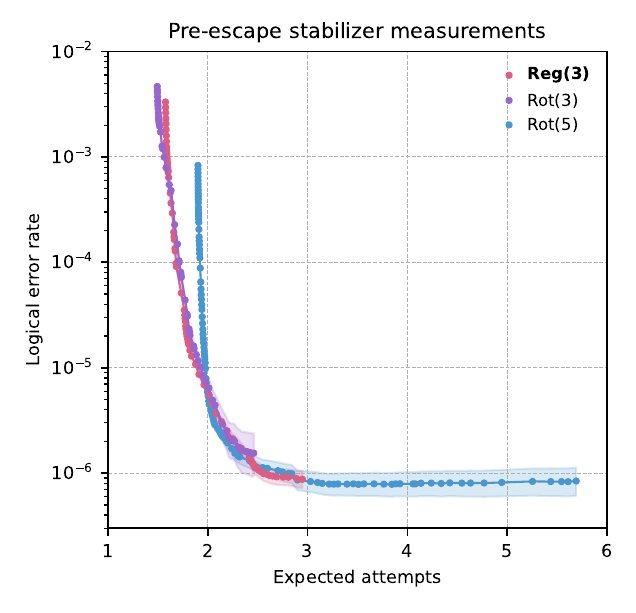}
                }; 
        \subfloat{\label{fig:ps3vary}}%
        \node at (6.5, 3.45) {\textbf{\textsf{(c)}}};
        \node at (8.9, -7.5) {
            \includegraphics[
                height = 0.4\textwidth,
                trim = 0 10 0 10, 
                clip
                ]{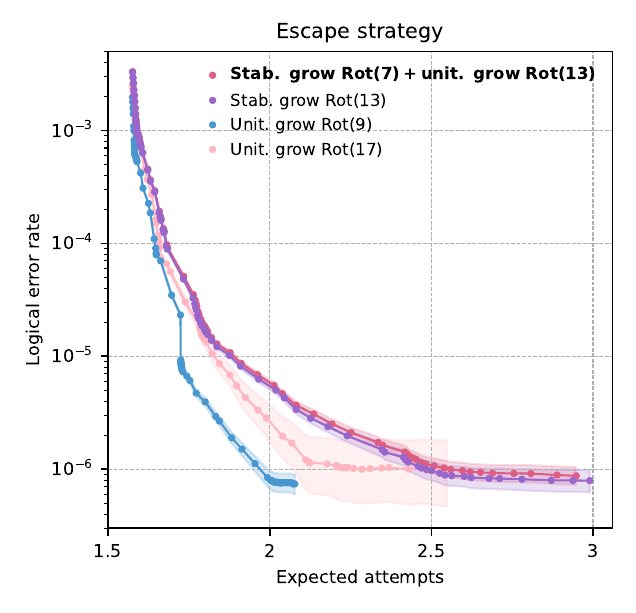}
                }; 
        \subfloat{\label{fig:escapevary}}%
        \node at (6.5, 3.45-7.5) {\textbf{\textsf{(d)}}};
    \end{tikzpicture}
    \caption{
        \textbf{Alternative choices for cultivation subroutines.} 
        Each panel shows the performance of the $\fd3$ $\ket{Y}$ state cultivation, but with one modified step. 
        The specific step that is modified is indicated by the title of the panels, and the modifications used are stated in the legend. Our `baseline' choice used in the main text is indicated in bold.
        A uniform depolarizing noise model with strength $p=0.001$ is used throughout, unless indicated otherwise in the legend. 
        }
\end{figure*}

\begin{enumerate}

    \item \textit{Hook injection~\cite{gidney2023cleaner,singh2022high}.} 
    This injection scheme consists of two rounds of stabilizer measurements. 
    In the first round, gates on $Z$-stabilizers along the diagonal are deliberately misordered such that a mid-cycle $T$ gate on an ancilla is converted into a logical operator. 
    We discard any shots with nontrivial measurement outcomes on fixed stabilizers --- or conflicting outcomes on unfixed stabilizers across these two rounds.
    For a uniform depolarizing noise model with strength $p$, the logical error rate of this scheme is $\frac{8}{15} p$. 
    Note that stabilizer-based injection schemes such as this require a subsequent measurement-based-fixup operation to ensure that the codespace is the +1 eigenspace of all stabilizers.
    
    \item \textit{Lao-Criger injection~\cite{lao2022magic}.} 
    This scheme also uses two rounds of stabilizer measurements and thus has a similar spacetime overhead to hook injection.
    For a uniform depolarizing noise model with strength $p$, the logical error rate of this scheme is $\frac{34}{15} p$. 
    This is higher than hook injection, therefore we do not consider this injection scheme further in this work.
    
    \item \textit{Unitary encoder + stabilizer measurements.} 
    Using a unitary encoding circuit should reduce the spacetime volume of injection.  
    Unlike optimizations made in Ref.~\cite{gidney2024magic}, here we consider a na\"ive unitary encoder with the physical $\ket{T}$ state as a part of the initial product state. 
    Note that unitary circuits intrinsically allow weight-1 errors to spread to weight-2 along a logical. 
    Consequently, they generally require a round of stabilizer measurements to detect and post-select out these mechanisms.
    
    \item \textit{`Optimized' unitary encoder + subset of stabilizer measurements.} 
    It is possible to design an optimized unitary encoding circuit that minimizes the spread of damaging errors. 
    This has been done in the quantum toolkit of Ref.~\cite{mqt2024}. 
    Using this package, we obtain an optimized unitary encoder that requires us to measure only three out of the eight stabilizers of $\rot3$ in order to detect weight-1 errors that spread along the logical.
    This further reduces the spacetime volume for injection compared to a na\"ive unitary encoder.
    This injection scheme is shown in~\cref{fig:optunit-f3} and discussed in~\cref{sec:nonlocal}.
    \item \textit{`Folding' unitary encoder + stabilizer measurements.} 
    Other magic state cultivation schemes use a unitary encoding circuit that first prepares a logical Pauli state, then `folds' in the logical operator, rotates it to the $\ket{\bar{H}_{XY}}$ state, and then folds it back out~\cite{gidney2024magic, claes2025cultivatingtstatessurface, vaknin2025magic}.
    This preparation scheme reduces the number of undetectable logical error mechanisms, but has a higher overall circuit depth. Stabilizer measurements are also required in this case to detect data qubit errors.
    
\end{enumerate}

\Cref{fig:injvary} provides a benchmark of the different injection schemes described above (excluding Lao-Criger injection).
{We plot the performance of $\fd3$ cultivation in the presence of uniform depolarizing noise with different strengths $p$.}
{We observe a consistent hierarchy of performance.}
For example, for all $p$, hook injection has the highest initial expected attempts, and the optimized unitary encoder has the lowest.

\begin{table}
    \centering
    \begin{tabular}{|l|c|c|}
        \hline
        \textbf{Injection scheme} & \textbf{Min. expected} & \textbf{Min. logical} \\
        & \textbf{attempts} & \textbf{error rate} \\
        \hline
        Hook injection & 1.62 & $8 \times 10^{-7} $\\
        Basic unitary enc. & 1.49 & $9 \times 10^{-7} $ \\
        Optimized unitary enc. & 1.46 & $13 \times 10^{-7} $ \\
        Folding unitary enc. & 1.54 & $7 \times 10^{-7} $ \\
        \hline
    \end{tabular}
    \caption{
        \textbf{Comparison of injection schemes} under uniform depolarizing noise of strength $p=0.001$.
        Quantified by the minimum expected attempts, and the error floors at full gap-based postselection. 
        Logical error rates have an uncertainty of roughly $1 \times 10^{-7}$.
        }
    \label{tab:floors}
\end{table}

{\Cref{tab:floors} summarizes the relative performance of each injection scheme at $p = 0.001$.}
We focus on two quantities: the number of initial expected attempts and the lowest logical error rate obtained when fully post-selecting on the complementary-gap.
We note that that, within error bars, the folding unitary encoding circuit reaches a logical error rate of $7 \times 10^{-7}$.
This is the required value provided in Ref.~\cite{zhou_resource_2025} for factoring of large integers. 
However, it has a higher initial expected attempts compared to other unitary encoders.

The na{\"ive} unitary encoding circuit has a balanced tradeoff between logical error rate and expected attempts. 
Therefore we use it for our simulations in the main text.
This choice may be amended based on the constraints of a given architecture, along with logical error rates required of the given application.
An injection scheme that achieves optimal performance remains the subject of further investigation.
Finally, we note that unitary-based injection schemes do not require mid-cultivation classical feedback.
As such, they may be more attractive to physical architectures where measurement and subsequent classical feedback is slow.

\subsection{Logical measurement}\label{app:logical-h-msmt}

In this section, we discuss the various choices related to measuring the $\bar{H}\XY$ operator after it has been injected.
It is important to perform this check efficiently while maintaining the fault distance of the protocol.

In this work, logical checks are performed with ancillary GHZ states. 
The size of the ancilla system
can impact the fault distance,
as well as the expected spacetime volume. 
For example, using only a single ancilla qubit for logical measurements is insufficient as a single error on it may propagate to multiple errors on data qubits, leading to logical errors. 
On the other hand, an ancillary GHZ($n$) state permits fault-tolerant logical measurements in an $n$-qubit code because any ancilla error does not propagate to more than one data qubit. 
However, this requires an unacceptably large spacetime volume for the fault-tolerant preparation of the large GHZ($n$) state.

Here we show that a GHZ(3) state is sufficient for measuring the $\bar{H}\XY$ operator of $\reg3$ with a fault distance 3. 
Similarly, GHZ(5) states are sufficient for $\reg{5}$ with a fault distance 5.

We first prove that in the absence of errors, a GHZ state of any size can be used to measure $\bar{H}\XY$. 
From~\cref{eq: HXY details}, we have that a $\bar{H}\XY$ consists of single qubit gates $UZU\dg$ for some $U$ on diagonal qubits and CZ gates on each pair of mirrored qubits. 
A controlled-$\bar{H}\XY$ gate therefore requires one CZ gate for each diagonal qubit and one CCZ gate for each pair of mirrored qubits. When an ancilla system is prepared in an $l$-qubit  $\GHZ{l}$ state stabilized by $X^{\otimes l}$, these CZ and CCZ gates can be implemented with their control on any of the ancilla qubits. This holds because, after all CZ and CCZ gates are applied, the stabilizer becomes 
\begin{equation}
    X^{\otimes l} \otimes \left(\prod_{i=0}^{2d-2}Z_{D_i}\right) \otimes \left( \prod_{q\in\Delta}CZ_{q, \tau(q)} \right),
\end{equation}
which is independent of the particular choice of control qubits of each gate. Finally, by applying a GHZ decoding circuit on the ancilla qubits, $X^{\otimes l}$ is mapped to a single-qubit $X$ on one ancilla qubit. Measuring this in the $X$ basis yields the logical measurement outcome of $\bar{H}\XY$.

\subsubsection{Data qubit partitions}

To achieve a desired fault distance $\mathsf{f}$, the logical measurement circuit has to be designed such that any physical error with weight less than $\mathsf{f}$ does not result in an undetectable logical error. 
This requires (i) a sufficiently large ancilla GHZ state $l \geq \mathsf{f}$, so that each ancilla qubit couples to only a limited number of data qubits, and (ii) a carefully chosen assignment of ancilla-data couplings, which prevents harmful correlated errors on the data. 
We present this assignment as a \textit{partition} of the data qubits into $l$ disjoint subsets, with each subset coupled to a distinct qubit of an ancillary $\GHZ{l}$ state. 
These partitions are represented by different colors in~\cref{fig:hxy-check-f3,fig:hxy-check-f5}. 

The partitions are constructed such that no combination of $\lfloor \mathsf{f}/2 \rfloor$ subsets fully supports a logical operator. 
For example, in~\cref{fig:hxy-check-f3} no single partition is connected a pattern of data qubits that support a logical operator on $\reg3$.
Otherwise, an error on that partition's ancilla could propagate to a logical error on the data, and a subsequent error on the same ancilla could cancel the first one.
This combination of two errors 
may introduce an undetectable logical error, lowering the fault distance of the logical check circuit.
In addition, it is desirable that the partition balances the number of gates on each ancilla qubit, which enables a shallow circuit and helps mitigate idle errors. 

\subsubsection{GHZ size}

While a $\GHZ{3}$ ancilla is sufficient to achieve a logical measurement circuit with fault distance 3, we also compare its performance with that of a larger $\GHZ{5}$ ancilla in~\cref{fig:ghzvary}. 
Using a larger ancilla reduces the circuit depth required during logical measurements, but comes at the expense of increased spacetime volume to prepare
the ancilla fault-tolerantly. 
In particular, we must include a flag qubit to ensure the fault-tolerant preparation of $\GHZ{5}$.
The results in~\cref{fig:ghzvary} include shots discarded purely due to flagged $\GHZ{5}$ preparation, and thus may overestimate the true overall discard rate. 
Compared to the smaller $\GHZ{3}$, $\GHZ{5}$ offers little advantage in terms of the expected spacetime volume. 
Thus, we choose $\GHZ{3}$ in our protocol due to its lower overhead.

\subsubsection{`Double-check' vs. `single-check'}
\label{app:double-v-single-check}

\begin{figure}
    \centering
    \includegraphics[width=0.95\linewidth]{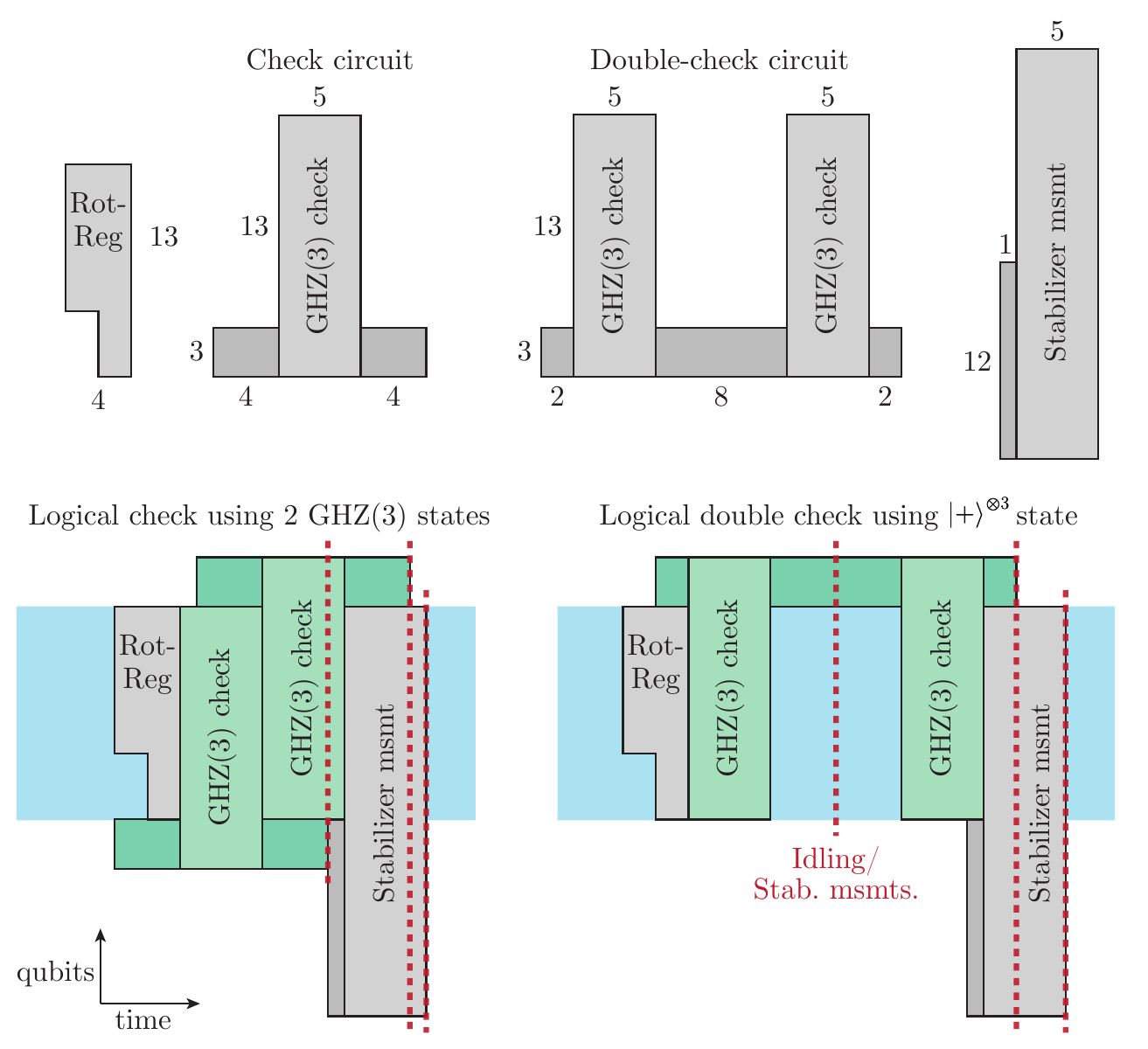}
    \caption{
        \textbf{Spacetime volume of logical checks}. 
        We show the space (vertical) and time (horizontal) cost of different cultivation steps.
        The top row shows the space and time costs of the preliminary gadgets that compose the logical checks.
        The lower left shows the spacetime volume of two sequential `single-checks' with pre-prepared GHZ(3) states, showing efficient pipelining.
        The lower right shows the spacetime volume of the `double-check' strategy of Ref.~\cite{gidney2024magic}.
        Red dashed lines represent postselection on measurement outcomes.
        }
    \label{fig:checkvsdc}
\end{figure}

Ref.~\cite{gidney2024magic} uses a `double-check' to measure a transversal Clifford operator twice using a single ancilla product state. 
This double-check provides two bits of information about the logical eigenvalue, along with error flags. 
Here, we simply choose to use two GHZ states to check this operator (once per GHZ state).
We obtain more efficient pipelining of operations since we are not required to wait for decoding and re-encoding of the ancillary system. 
Consequently, we bypass the cost of either having a longer idling time on the initial code during which errors can build up, or having to do more stabilizer measurements during this time.

\Cref{fig:checkvsdc} shows a spacetime volume comparison of these alternative check strategies. 
The prepared GHZ states allow more compact cultivation, albeit at the cost of using more qubits.

\subsubsection{{Higher fault distances}}

\Cref{fig:hxy-check-f5} shows a partition of a $\reg{5}$ code into five subsets that enables fault-tolerant $\bar{H}\XY$ measurements. 
For the partition to be valid, it must prevent any weight-4 physical error from propagating into an undetectable logical error on the data.
To guarantee $\fd5$, we impose the condition that the union of any two subsets must not contain the full support of a logical operator. 
Similar to the $\reg3$ code, the most dangerous scenario arises when two errors occur on two ancilla qubits, propagate to the data, and then two additional errors on the same ancillae cancel the earlier ones, rendering the error undetectable.

For higher fault distances, we must also increase the number of measurements of the fold-transversal $\bar{H}\XY$ operator. 
Specifically, we must measure $\bar{H}\XY$ at least $\mathsf{f}-1$ times for a fault distance $\mathsf{f}$. 
To see why, consider the following adversarial error:
\begin{enumerate}
    \item A single error during injection may cause a logical error, as the injection circuit is not fault-tolerant.
    \item A measurement error on the outcome of $\bar{H}\XY$ may hide this logical error, yielding $+1$ and leading to the belief that no error has occurred.
\end{enumerate}
If we only measure the logical operator $t < \mathsf{f}-1$ times, then we have a logical error on the code with only $t+1$ errors.
To ensure that logical errors only occur after $\mathsf{f}$ errors, must measure the logical operator $\mathsf{f}-1$ times. 
For $\fd5$, we can achieve this by doing two checks at $\reg3$, and two checks at $\reg{5}$. 
This design reduces expected spacetime volume compared to checking four times at $\reg{5}$.

\subsection{Pre-escape stabilizer measurement}\label{app:regps}

In this section, we discuss different options for extracting stabilizer information after the logical checks have been performed, but prior to the escape stage.
This is used for additional postselection in order
to achieve low error rates~\cite{chamberland2020very}. 
For example, in our $\fd3$ protocol a single round of stabilizer measurements is needed after the logical checks at $\reg3$.
However, is not strictly necessary to perform this check at $\reg3$, and unitary circuits may be used to transform to another code where stabilizer measurements improve performance. 

\Cref{fig:ps3vary} shows the performance of cultivation for three different choices of code in which to perform these `pre-escape' stabilizer measurements.
We consider the following options: (i) stabilizer measurements at $\reg3$ using a depth-4 circuit, (ii) stabilizer measurements at $\rot3$ using a depth-2 circuit, and (iii) growing to $\rot5$ and then measuring stabilizer using a depth-6. 
Measuring at $\reg3$ offers the best performance in terms of expected attempts and logical error rate.

\subsection{Escape strategy}\label{app:escapestrat}

In this section, we discuss different escape strategies.
The primary goal of the escape stage is to expand to a larger code of distance $d_2  > 2 \mathsf{f}$.
{This is done} as quickly as possible in order to preserve the error scaling achieved by cultivation.
A secondary goal is to collect stabilizer information from this larger code in order to post-select based on the complementary-gap.
Note that $d_2$ is the minimum code distance necessary to protect the magic state.
This is often going to be different to the final code distance $d\fin$ required for use in a quantum algorithm.
It most important to grow to $d_2$ rapidly; the subsequent growth strategy to $d\fin$ is done after all postselection.

Ref.~\cite{gidney2024magic} uses stabilizer measurement rounds to grow directly to $d\fin$. 
This stabilizer-based escape is expensive in terms of spacetime volume, since several gate and measurement layers at $d\fin$ are required.
Ref.~\cite{chen2025efficient} also uses stabilizer-based escape, but only to $d_2$, thus overlooking the possible overhead associated with growing to $d\fin$.

Recently, efficient unitary encoders have been proposed that enable conversion from small- to large-distance rotated surface codes in low depth~\cite{tsai2025unitary, higgott2021optimal}. 
While efficient in terms of spacetime volume, a unitary escape strategy directly to $d\fin$ does not by itself provide the stabilizer information necessary for gap-based-postselection. 
Nevertheless, we can examine its `theoretical' performance.

\Cref{fig:escapevary} provides a comparison of unitary-based and stabilizer-based escape strategies.
To benchmark purely-unitary escape, we use a perfect measurement round at the end of growth to compute a complementary-gap and obtain a logical error rate. 
Though this is unphysical due to the perfect round of measurement, unitary escape has both a lower spacetime overhead and a lower error floor than stabilizer-based escape.
This demonstrates the limitations imposed by escape.
We also remark that in a setting where measurements are noisier than gates, such as in neutral atoms, one may further benefit from using unitary escape instead of measurement-based escape.

In this work, we utilize a \textit{three-stage escape} strategy that is a combination of unitary- and stabilizer-based growth steps.
After the logical checks in $\mathsf{Reg}$, we unitarily grow to $\mathsf{Rot}$.
Then, we use stabilizer-based escape to grow to $\rot{d_2}$, where we post-select on the complementary gap.
Finally, we unitarily grow to $\rot{d\fin}$. 
In doing so, we avoid the large spacetime overhead of stabilizer measurement rounds at $d\fin$.

\section{Fault distance 5}\label{app:fd5}

In~\cref{sec:protocol} we presented a cultivation protocol for fault distance $\mathsf{f=3}$. 
Here, we extend our protocol to a larger fault distance $\mathsf{f=5}$. 
This variant of our
protocol reaches a lower logical error rate by further fault-tolerantly checking the $\bar{H}\XY$ eigenstate at $\reg{5}$.  
Similar to previous works, we find that a simple modification that changes our $\fd5$ scheme to $\fd 4$ achieves comparable logical error rate with lower overhead~\cite{gidney2024magic}. 

We begin by outlining our end-to-end $\mathsf{f=5}$ protocol. 
The initial stages 
are identical to that of $\mathsf{f=3}$.
Specifically, from injection up to the first two $H\XY$ checks at $\reg3$. 
Following this,
we add the following steps: 
\begin{enumerate}
    \item Unitarily grow to $\rot5$.
    \item Perform $r_1$ stabilizer measurement rounds.
    \item Unitarily grow to $\reg{5}$ and perform two additional logical $\bar{H}\XY$ checks using a $\GHZ{5}$ ancilla.
    \item Three-stage escape to $\rot{d\fin}$ which involves $r_2$ stabilizer measurement rounds at $\reg{5}$.
\end{enumerate}

\Cref{fig:LER} shows the logical error rate of $\fd5$ $\ket{Y}$ state cultivation under uniform depolarizing noise model with a strength $p = 0.001$.
Setting $(r_1, r_2) = (2,0)$ we can reach a logical error rate of $10^{-9}$ in fewer than 10 expected attempts.
This is an improvement over previous works that require approximately $20$ expected attempts in order to reach this regime.
In the following paragraphs, we re-iterate the protocol from end-to-end for clarity. 

\textit{Injection---}  
We use a unitary encoding circuit to inject the $\ket{\bar{H}\XY}$ state into $\rot3$. 
This is followed by a round of stabilizer measurements.
We post-select on any non-trivial stabilizer measurement outcomes.

\textit{Cultivation---} 
We use a unitary circuit that to transform to $\reg3$ and measure $\bar{H}\XY$ twice with a $\mathrm{GHZ}(3)$ ancilla. 
We then grow to $\rot5$ and perform $r_1$ rounds of stabilizer measurements.

Next, we grow to $\reg{5}$ and measure $\bar{H}\XY$ twice again, except this time with a $\mathrm{GHZ}(5)$ ancilla.
We show the circuit for this check in~\cref{fig:hxy-check-f5} (and the equivalent $\bar{Y}$ check in~\cref{fig:f5-Y-check}). 
The $\mathrm{GHZ}(5)$ ancilla uses
an extra flag qubit such that no single physical error leads to a weight-2 error on the $\mathrm{GHZ}(5)$ state. 

At this stage, we may choose to measure stabilizers of $\reg{5}$ surface code $r_2$ times to further lower the LER before escape. 
Through numerical search for the minimum-weight error that causes a logical error, we observe that $r_1 \geq 3$ and $r_2\ge1$ are necessary to achieve $\fd5$ scaling. 
This is because high-weight errors may appear in $\rot5$ arising from a single error before growth at $\reg3$. 
These errors can remain undetected by all subsequent $\bar{H}\XY$ checks in $\rot5$. 
They may then grow to undetectable logical operators in future steps, reducing the fault distance when $r_1 < 3$.

\textit{Escape---} 
We use a three-stage escape strategy.
First, we unitarily grow to $\rot 9$.
Next, we use stabilizer-based growth to $\rot{d_\mathrm{mid}}$, using $r_\mathrm{mid}$ rounds of stabilizer information for gap-based postselection.
Finally, we unitarily grow to $\rot{d\fin}$.

For our numerical simulations, we perform $r_\mathrm{mid}=5$ rounds of stabilizer measurements on an intermediate code of distance $d_\mathrm{mid} = 11$.
We choose a final code distance $d\fin=15$ to be consistent with the choice in Ref.~\cite{gidney2024magic}. 
For the third stage of escape we use the unitary growth circuit proposed in Ref.~\cite{claes2025lowerdepthlocalencodingcircuits}.
If we were to use the same unitary growth circuit as in the $\fd3$ protocol, the code distance would double in a single step, taking $d_\mathrm{mid} = 11$ to $21$, which exceeds our target final size. 

\begin{figure}
    \centering
    \includegraphics[
        width = \linewidth,
        trim = 0 10 0 10, 
        clip
        ]{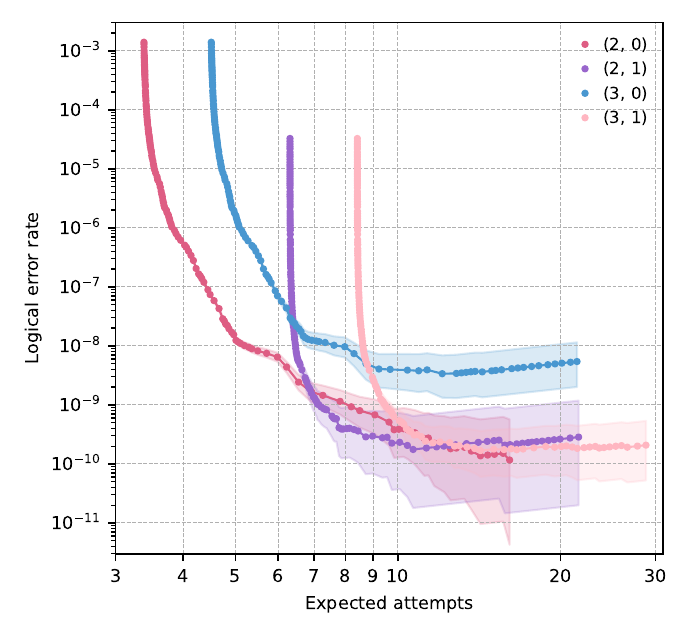}
    \caption{
        \textbf{Varying stabilizer measurements $(r_1, r_2)$} for fault distance $\fd5$ cultivation.
        Here, $r_1$ is the number of stabilizer measurement rounds at $\rot5$, and $r_2$ the number of rounds at $\reg{5}$. 
        }
    \label{fig:f5-gap-plot}
\end{figure}

\Cref{fig:f5-gap-plot} shows the performance of $\fd5$ cultivation for different numbers of stabilizer measurement rounds $(r_1, r_2)$.
We choose {$(2,0)$} for our protocol.
It has comparable error floor to {$(2,1)$ and $(3,1)$ schemes}, but lower overhead as it involves less stabilizer measurement rounds. 
Through exhaustive numerical search, the comparable error floors of these alternatives is attributed to an extremely low number of weight-4 logical error mechanisms in $(2, 0)$. 
We also disregard the $(3,0)$ scheme, because it has a higher error floor of compared to the $(2,0)$ scheme. 
Since both $(3,0)$ and $(2,0)$ have a fault distance of $4$, we suspect that the increased error floor is due to a greater number of error mechanisms introduced by the additional $\rot3$ check at $(3,0)$. 
Note that all data shown here includes shots discarded due to flags during $\GHZ{5}$ state preparation.
Thus, it may overestimate the number of shots required for cultivation using a set of pre-prepared $\GHZ{5}$ states by about $\sim 1\%$.

\section{Handoff simulation}\label{app:handoff}

In this section, we describe our `handoff' simulation that combines both state vector and stabilizer simulators. 
We use this technique to simulate the logical error rate of $\ket{\bar{H}\XY}$ magic state cultivation scheme.
This is to be contrasted with previous work that instead simulates $\ket{\bar{Y}}$ state cultivation as a proxy~\cite{gidney2024magic,vaknin2025magic,chen2025efficient,claes2025cultivatingtstatessurface}.
Our handoff simulation directly samples noisy logical $\ket{\bar{H}\XY}$ states produced by the full protocol without approximations, while remaining tractable for fault distance $\fd3$. 

We achieve this by splitting our circuit into two parts, which are simulated by a state vector simulator and a stabilizer simulator, respectively. 
The state vector simulation starts with the injection of $\ket{\bar{H}\XY}$ and ends after the completion of the $\bar{H}\XY$ measurements.
This part of the protocol contains all the physical non-Clifford operations, and is supported on a relatively small number of qubits.
The stabilizer simulation starts from the stabilizer measurements after the $\bar{H}\XY$ measurement and ends after the completion of the escape stage. 
This part of the circuit contains only Clifford gates, and can be simulated efficiently with a stabilizer simulator such as \texttt{Stim}. 

In~\cref{sec:handoff-steps} we describe the steps of the handoff simulation, and how we combine the state vector and stabilizer segments.
Then, in~\cref{sec:handoff-proof} we prove that the simulation exactly samples the noisy $\ket{\bar{H}\XY}$ states.
Finally, in~\cref{sec:handoff-results} we discuss the numerical results and compare the cultivation of $\ket{\bar{Y}}$ states vs. $\ket{\bar{H}\XY}$ states.
We remark here that our approach can be considered as an adaptation of the splitting technique from Ref.~\cite{lima_clifford_2025} to magic state cultivation.

\subsection{Procedure}\label{sec:handoff-steps}

\newcommand{\A}{_a}
\newcommand{\B}{_b}

In this section, we describe the steps of our handoff simulation  for fault distance $\fd3$ cultivation.

Each step involves noisy operations unless specified otherwise.
The details of our cultivation protocol are summarized in~\cref{fig:summary} and described in~\cref{sec:protocol}.

\subsubsection{State vector simulation}

First, we use a state vector simulation to extract the noisy state of $\reg3$ immediately following $\bar{H}\XY$ checks.
This is achieved by means of a noiseless round of stabilizer measurements.
These are \textit{not} the stabilizer measurements used for postselection.
Instead, they are a means of extracting the logical state of the code, as well as the state of the data qubits.
The simulation proceeds as follows.

\begin{enumerate}

    \item  
    Simulate the cultivation protocol up to and including the $\bar H\XY$ checks.
    Post-select on any nontrivial measurement outcomes.
    Namely, the outcomes of the stabilizer measurements during injection and the $\bar H\XY$ checks during cultivation.

    \item
    Obtain the state vector $\ket{\psi}\in \mathbb{C}^{2^{13}}$ that describes the thirteen data qubits of $\reg3$.
    
    \item  
    Perform a round of noiseless stabilizer measurements and obtain the bit string $s\in\{0,1\}^{12}$ corresponding to the syndromes.

    \item 
    Obtain the (normalized) state vector $\ket{\psi_s}\in \mathbb{C}^{2^{13}}$ by projecting $\ket{\psi}$ onto the subspace defined by $s$.
    This is the state of the data qubits after measuring the ancilla qubits and obtaining the measurement outcomes $s$.
    
    \item 
    Construct a \textit{Pauli destabilizer} $D_s$ by multiplying strings of physical $X$ and $Z$ operators that cancel all nontrivial syndromes in $s$ by connecting each nontrivial syndrome to the bottom or right boundaries of the code.

    \item 
    Rewrite $\ket{\psi_s}$ in the form of a logical state vector multiplied by the destabilizer $D_s$:
    \begin{equation}
        \ket{\psi_s} = D_s(\alpha\ket{0\A} + \beta\ket{1\A}),
    \end{equation}
    where $\ket{0\A}$ and $\ket{1\A}$ are the logical basis states of $\reg3$ and the coefficients are given by
    \begin{align}
        \alpha = \braket{0\A|D_s\dg|\psi_s}, 
        \label{eq:logical-alpha}
        \\
        \beta = \braket{1\A|D_s\dg|\psi_s}.
        \label{eq:logical-beta}
    \end{align}
    
\end{enumerate}

\textit{Remark 1:}  In practice, steps 2 and 3 above can be condensed.
This is achieved by applying the inverse of the unitary encoding circuit $U\A$. 
Applying $U\A\dg$ maps a representative of the logical operators $\bar{X}\A$ and $\bar{Z}\A$ onto the $X$ and $Z$ operators of a single data qubit.
Simultaneously, it maps the product of stabilizer generators onto single-qubit $Z$ operators on the twelve other data qubits. 
Therefore, with appropriate post-processing, the syndrome bit string $s$ and the logical coefficients $\alpha$ and $\beta$ can be determined 
by measuring the thirteen data qubits.

\textit{Remark 2:} 
The noiseless stabilizer measurement outcomes $s$ are used only for the purposes of handoff; 
we do not use $s$ for decoding.
        
\subsubsection{Stabilizer simulation}

At this point, we handoff the results to a stabilizer simulator.
The purpose of the stabilizer simulator is to sample logical Pauli errors that may occur after the $\bar{H}\XY$ checks.
As such, the stabilizer simulator only takes the destabilizer $D_s$ as input from the state-vector simulator.
The logical state vector coefficients $\alpha$ and $\beta$ are stored in order to reconstruct the noisy magic state at the end of the protocol.
The simulation proceeds as follows.

\begin{enumerate}

    \item 
    Prepare the logical state $\ket{0\A}\in\mathbb{C}^{2^{13}}$ of the $\reg3$ code using a noiseless unitary encoding circuit~\footnote{Any logical Pauli state is sufficient here. The choice is irrelevant because further measurements and corrections are independent of the encoded logical state.}. 
    Then, apply the destabilizer $D_s$ obtained from the previous state vector simulation.
    
    \item 
    Proceed with the remainder of the protocol.
    Perform a noisy round of stabilizer measurements of $\reg3$ and post-select for the trivial syndrome.
    If successful, proceed with escape, obtaining measurement outcomes $g$ along the way.
    
    \item 
    Terminate the simulation at the end of MSC with a round of noiseless stabilizer measurements on $\rot{d\fin}$ and obtain the measurement outcomes $q$.
    
    \item 
    Propagate $D_s$ and the physical Pauli errors that were sampled in the stabilizer simulation to the end of line on each physical qubit in software. 

    \item 
    Use two bits $x$ and $z$ to record whether the propagated error anticommutes with the logical operators of $\rot{d\fin}$, denoted $\bar{X}\B$ and $\bar{Z}\B$.
    Here, $\mathtt{True}$ is recorded as 1.

    \item 
    Decode using the syndrome information obtained from the stabilizer measurements made during escape, $g$ and $q$.
    Recall that this is another point of postselection, depending on the confidence of the gap-based decoder.
    From this, we obtain the correction operator $R$.
    
    \item 
    Store two bits $x_R$ and $z_R$ to record whether the recovery $R$ anticommutes with $\bar{X}\B$ and $\bar{Z}\B$.
    Again, with $\mathtt{True}$ being recorded as 1.
    
\end{enumerate}
    
\subsubsection{Magic state reconstruction}

Finally, we combine the two simulations.
The noisy magic state sampled by the handoff simulation is
\begin{align}
    \ket{\tilde{H}\XY} 
    = 
    \bar{X}\B^{z+z_R} \bar{Z}\B^{x+x_R}\left(\alpha\ket{0\B} + \beta\ket{1\B}\right)
    , 
    \label{eq:H_XY-sample}
\end{align}
where $\ket{0\B}$ and $\ket{1\B}$ are the logical basis states of $\rot{d\fin}$.
Recall that $\alpha$ and $\beta$ are obtained from the state vector simulation, while the bits $x$, $z$, $x_R$, and $z_R$ are obtained from the stabilizer simulation.
We declare whether a logical error has occurred by comparing the noisy magic state to the ideal magic state
$ \ket{\bar{H}\XY} = (\ket{0\B} + e^{i\pi/4}\ket{1\B})/\sqrt{2}$.
A logical error has occurred when we find that
\begin{equation}
    \big|\! \braket{\bar{H}\XY|\tilde{H}\XY} \!\big|^2 < 1.
\end{equation}

\subsection{Working principle}\label{sec:handoff-proof}

In this section we prove that our handoff simulation exactly samples the noisy $\ket{H\XY}$ states without approximation.
First, we show that inserting a round of `fictitious' and noiseless stabilizer measurements after the $\bar{H}\XY$ checks on $\reg3$
does not disturb the outputs of the noisy cultivation circuit. 
Intuitively, this is because any state vector collapse caused by the noiseless measurements happens regardless due to the \textit{actual} noisy stabilizer measurements performed shortly thereafter.

Consider the state $\ket{\psi}$ of the data qubits following the logical checks.
It is convenient to rewrite this state as
\begin{equation}
    {\ket{\psi} = \sum_s c_s \ket{\psi_s},}
\end{equation}
where $\ket{\psi_s}$ is the state after projection onto the subspace defined by the syndrome bit string $s$, and $c_s$ is a complex coefficient.
The noisy syndrome extraction circuit may be expressed in the form
\begin{equation}
    \tilde{C} = P\dat P\anc C,
\end{equation}
where $C$ is the noiseless circuit, and $P\dat$ and $P\anc$ are Pauli errors on the data and ancilla qubits, respectively.
The noiseless syndrome extraction circuit $C$ maps the syndromes $s$ onto the ancilla qubits:
\begin{equation}\label{eq:stab-msmt}
    {C (\ket{\psi} \otimes \ket{0}^{\otimes n\anc})
    =
    \sum_s c_s (\ket{\psi_s} \otimes \ket{s})
    .}
\end{equation}

The noisy circuit $\tilde{C}$ has the following action on the data and ancilla qubits:
\begin{equation}\label{eq:after-stab-msmt} 
    \begin{aligned}
        \tilde{C} (\ket{\psi} \otimes \ket{0}^{\otimes n\anc}) 
        &= P\dat P\anc \sum_s c_s (\ket{\psi_s} \otimes \ket{s}) \\
        &= \sum_s c_s (P\dat \ket{\psi_s} \otimes P\anc \ket{s})
        .
    \end{aligned}
\end{equation}
Upon measurements of the ancilla qubits, we obtain the state $P\dat\ket{\psi_s} \otimes P\anc\ket{s}$ with probability $|c_s|^2$.

Now we will show
that the handoff circuit with noiseless stabilizer measurements preserves the same state and statistics as~\cref{eq:after-stab-msmt}.
As before, the initial state is $\ket{\psi} \otimes \ket{0}^{\!\otimes n\anc}$. 
First, we perform the round of noiseless stabilizer measurements that are used for handoff.
From~\cref{eq:stab-msmt}, measurement of the ancilla qubits projects onto the state $\ket{\psi_s} \otimes \ket{s}$ with probability $|c_s|^2$.
Then, we apply the noisy syndrome extraction circuit $\tilde{C}$:
\begin{equation}
    \begin{aligned}
        \tilde{C} (\ket{\psi_s} \otimes \ket{0}^{\otimes n\anc}) 
        &= P\dat P\anc (\ket{\psi_s} \otimes \ket{s}) \\
        &= P\dat \ket{\psi_s} \otimes P\anc \ket{s}.
        \label{eq:handoff-after-stabmsmt}
    \end{aligned}
\end{equation}
Which is the same state as~\cref{eq:after-stab-msmt}.
The noiseless circuit $C$, followed by ancilla measurement, and then the noisy circuit $\tilde{C}$ results in the same state $P\dat\ket{\psi_s} \otimes P\anc\ket{s}$ with the same probability $|c_s|^2$.
Therefore, the noiseless stabilizer measurements do not disturb the outputs of the noisy stabilizer measurements in the cultivation stage, nor the outputs of the rest of the circuit. 

After the stabilizer measurements we are left with a state $\ket{\psi_s}$ that
is confined to a two-dimensional subspace defined by the syndrome bit string $s$.
A basis for this two-dimensional subspace is $D_s\ket{0\A}$ and $D_s\ket{1\A}$, where $D_s$ is the Pauli destabilizer defined in step 5 of the state vector simulation. 

We can write the state as
\begin{align}
    \ket{\psi_s} 
    &= D_s(\alpha \ket{0\A} + \beta \ket{1\A}), 
    \label{eq:psi_s-expression}
\end{align}
where $\alpha$ and $\beta$ are given by~\cref{eq:logical-alpha,eq:logical-beta}.
In this way, the syndrome bit string $s$ and the complex coefficients $\alpha$ and $\beta$ completely determine state $\ket{\psi_s}$.

We now wish to simulate the evolution $\ket{\psi_s}$ through the rest of the cultivation protocol. 
While $\ket{\psi_s}$ is not guaranteed to be a stabilizer state, we will see that the measurement outcome statistics of the stabilizer simulation
are unchanged if the state $\ket{\psi_s}$ is replaced by a stabilizer state $D_s\ket{0\A}$. 
This is because the remaining Clifford circuit is constructed to measure either stabilizers or operators that anticommute with stabilizers, and not logical operators. 
The sampling of measurement outcomes can be achieved in the stabilizer simulator by preparing $D_s\ket{0\A}$, or indeed any logical Pauli state.

\begin{figure*}\label{fig:app_hxy_compare}
    \centering
    \begin{tikzpicture}
        \node at (0, 0) {
            \includegraphics[
                width = 0.48\linewidth,
                trim = 0 10 0 10, 
                clip
                ]{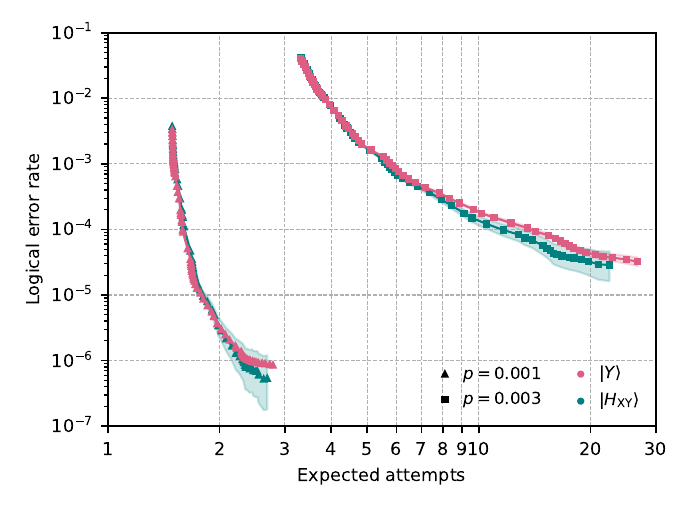}
            \subfloat{\label{fig:app_hxy}}%
            \includegraphics[
                width = 0.47\linewidth,
                trim = 0 10 0 10, 
                clip
                ]{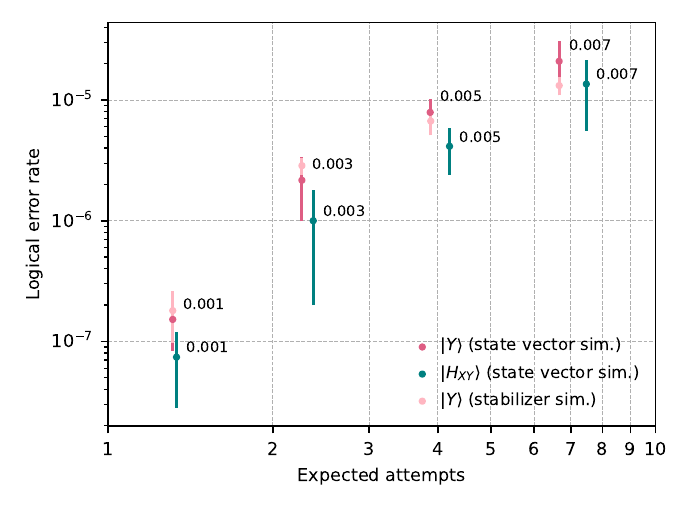}
            \subfloat{\label{fig:app_hxy_conly}}%
                };
        \node at (-7.05, 3.025) {\textbf{\textsf{(a)}}};
        \node at (1.65, 3.025) {\textbf{\textsf{(b)}}};
    \end{tikzpicture}
    \caption{
        \textbf{Comparing \boldsymbol{$\ket{\bar{Y}}$} and \boldsymbol{$\ket{\bar{H}\XY}$} state cultivation} (a) including the escape stage and (b) excluding the escape stage. 
        Simulations use the fault distance $\fd3$ protocol, with uniform depolarizing noise of strength $p$.
        When we exclude escape, the two states have different logical error rates.
        }
\end{figure*}

The final noisy magic state is obtained from~\cref{eq:H_XY-sample}.
Note that this depends on the logical coefficients $\alpha$ and $\beta$ (extracted from the state vector simulation), and the bits $x$, $z$, $x_R$ and $z_R$ (extracted from the stabilizer simulation).
Only the syndrome bit string $s$ from $\reg3$ is passed between the two simulations, in the form of the destabilizer $D_s$.
Finally, we point out that if the non-Clifford gates transform Pauli errors in a simple way, then magic state preparation can be exactly simulated on a stabilizer simulator.
Specifically, by rewriting the magic state density matrix into a linear combination of the density matrices of pure stabilizer states~\cite{daguerre_code_2025}.

\subsection{{\texorpdfstring{$\ket{\bar{Y}}$}{Y} vs. \texorpdfstring{$\ket{\bar{H}\XY}$}{H} cultivation}}\label{sec:handoff-results}

Here, we further compare $\ket{\bar{Y}}$ and $\ket{\bar{H}\XY}$ state cultivation for our $\fd3$ scheme.
\Cref{fig:app_hxy} shows the logical error rate of cultivation for the two states under uniform depolarizing noise of different strengths $p$.
We find that the relationship between the LER of cultivation and the expected attempts is very similar for the two different states.

In previous work, a truncated state vector simulation was used to estimate the relative performance of Clifford and non-Clifford state cultivation~\cite{gidney2024magic}.
In these simulations, the injection and cultivation stages are truncated by a perfect round of stabilizer measurements and there is no escape stage.
\Cref{fig:app_hxy_conly} shows the results of such simulations for our protocol.
We find that $\ket{\bar{Y}}$ and $\ket{\bar{H}\XY}$ state cultivation performs differently in this setting. 
Additionally, we observe that, at higher $p$ values magic state cultivation has a lower logical error rate and correspondingly more expected attempts than the Clifford state. 
This behavior is in contradiction to what is seen in previous work~\footnote{Fig. 13 of Ref.~\cite{gidney2024magic}}.
To verify that our truncated state vector simulation is performing reliably, we have also used a stabilizer simulator for the $\ket{\bar{Y}}$ state. 
This stabilizer simulation agrees with the state vector simulation.

Taken as a whole, these simulations suggest that the logical error rate of the protocol is dominated by error mechanisms during escape. 
It follows that the truncated simulations are thus not a reliable indicator of the true end-to-end performance of cultivation protocols.

dominated by error mechanisms during escape.

\section{Numerical simulations}\label{app:sims}

In this section, we present details as to how we obtain and plot the data we show in the main text. 
Note that sample code and circuits are included at Ref.~\cite{gitrepo}. 
Data was generated using \texttt{sinter} and \texttt{PyMatching}~\cite{sparse25higgott, gidney_sinter}.

\subsection{Complementary gap plots}

\textit{Data from other schemes ---}
We obtained the data for~\cref{fig:e2e} from Ref.~\cite{gidney2024magic, chen2025efficient}'s respective Zenodo repositories. 
Note that Ref.~\cite{chen2025efficient} uses 2D soft outputs in terms of the parameters $(\phi_{\mathbb{RP}^2}, \phi_{bd})$ instead of the 1D complementary gap. 
For plotting purposes we convert the data therein to a 1D gap using the function $G = \phi_{\mathbb{RP}^2}+  0.2 \phi_{bd} $. 
To the best of our knowledge, this converted data is near-indistinguishable from the optimized points chosen from the 2D space shown in the original manuscript.

\textit{Error bars ---}
The error bars shown are calculated using an inbuilt \texttt{sinter} function, highlighting a region within a factor of 1000 of the maximum likelihood. 
Note that for handoff we performed $50$ fewer state vector shots, and oversampled the Clifford circuits on these outputs.

\textit{Final, perfect measurements ---}
In an actual MSC protocol, perfect syndrome measurement is not present, and thus cannot contribute to the computation of the complementary gap $G$. However, our simulations include a final perfect round of measurements. 
In order to verify that our results do not rely on this noiseless information for computation of the gap, we tested different numbers of stabilizer measurement rounds during escape. 
Note that the complementary-gap plots for only unitary growth-based escape solely rely on this perfect round and are thus unphysical.

\textit{Baseline scheme used in ~\cref{fig:ps3vary,fig:escapevary} ---}
In these figures, the state preparation was achieved by means of hook injection, instead of a unitary encoder used in other simulations. 
Due to the modularity of MSC, we expect that the relative hierarchy of performance between the curves should not change if we were to use a unitary encoder. 

\subsection{Noise models}

\begin{table}
    \centering
    \begin{tabular}{|r|l|l|l|}
        \hline
        \textbf{Abbreviation}
        &SD6
        &PM1 or PM5
        \\\hline
        \textbf{Name}
            & \begin{tabular}{@{}l@{}}Standard\\Depolarizing\end{tabular}
            & \begin{tabular}{@{}l@{}}Physically\\Motivated\end{tabular}

        \\\hline
        \textbf{Noisy Gateset}
            &\noindent\begin{tabular}{@{}l@{}}
                $\text{CX}(p)$\\
                $\text{CY}(p)$\\
                $\text{CCZ}(p)$\\
                $\text{AnyClifford}_1(p)$\\
                $\text{Init}_Z(p)$\\
                $M_Z(p)$\\
                $\text{Idle}(p)$\\
                
            \end{tabular}
            &\begin{tabular}{@{}l@{}}
                \vspace{-0.25cm}
                {} \\
                $\text{CZ}_{\text{local}}(p)$\\
                $\text{CZ}_{\text{non-local}}(p \text{ or }5p)$\\
                $\text{CCZ}(5p)$\\
                $\text{AnyClifford}_1(p/10)$\\
                $\text{Init}_Z(p)$\\
                $M_Z(p)$\\
                $\text{Idle}(0)$\\
                
            \end{tabular}
        \\\hline
    \end{tabular}
    \caption{
        \textbf{Noise models.} 
        Noise channels are applied after the unitary operation or reset, or before measurement.
    }
    \label{tbl:models}
\end{table}

In~\cref{tbl:models}, we show the noise models used in this work. 
A noisy unitary operation on $n$ qubits is associated with the relevant depolarizing Pauli channel $\{I,X,Y,Z\}^{\otimes n} \backslash I^{\otimes n}$ occurring with the stated probability. 
For measurement and reset in the $Z$ basis, we apply a probabilistic $X$ error before and after the operations, respectively.

\textit{Note added:} In a previous version of this manuscript, reported $\fd{5}$ data incorrectly did not have idling errors at a subset of circuit locations. Specifically, idles were excluded on $\reg{3}$ qubits that did not participate in the logical CY check for the duration of the check. These are the same qubits absent in $\rot{3}$. This has now been corrected, leading to a slight increase in the number of expected attempts for $\fd{5}$ data, and a higher error floor for the $(3,0)$ scheme in \cref{fig:f5-gap-plot}.

\subsection{Spacetime volume} \label{app:spacetime volume}

We calculate the expected spacetime volume of protocols in the following way.
For a cultivation circuit, let $M$ denote the number of postselection stages which divide the circuit into $M+1$ segments labeled $C_i$.
Each segment $C_i$ has a spacetime volume $V_i$, defined as the product of the number of active qubits and the number of gate steps (circuit depth). 
Let $f_i$ denote the fraction of shots that proceed through the first $i$ postselection stages, with $f_0=1$ by definition. At the end, a fraction $f_M$ of shots successfully passes all postselection stages, so on average the protocol has to be repeated $1/f_M$ times to obtain a successful shot. 

The expected spacetime volume per successful shot is
$V = \frac{1}{f_M} \sum_{i=0}^M f_i V_i$.
Each term $f_i V_i$ accounts for the expected cost of executing segment $C_i$, weighted by the fraction of shots that reaches it. 
The prefactor $1/f_M$ accounts for the fact that multiple attempts are typically required before obtaining a kept shot. 
Note that when there is no circuit required after the last postselection stage, such as complementary-gap-based postselection in Refs.~\cite{gidney2024magic, chen2025efficient, claes2025cultivatingtstatessurface, vaknin2025magic}, $V_M=0$. 

To make a fair comparison of the expected spacetime volume at a fixed final code distance in~\cref{fig:STO}, we estimate the volume of the protocols in Refs.~\cite{gidney2024magic, chen2025efficient, claes2025cultivatingtstatessurface, vaknin2025magic} at $d\fin=13$ (for $\mathsf{f}=3$) and $d\fin=15$ (for $\mathsf{f}=5$). 
All stages prior to the final escape are independent of the final distance, so only the escape stage requires adjustment. 
In the escape stage, the spacetime volumes of the escape circuits themselves can be straightforwardly rescaled to $d\fin$. 
For the final success rates, we note that in our protocol the tradeoff between logical error rate and expected attempts, shown in~\cref{fig:LER}, does not depend strongly on the final code distance. 
Guided by this observation, we adopt the reported curves from Refs.~{\cite{gidney2024magic, chen2025efficient, claes2025cultivatingtstatessurface, vaknin2025magic} with different final distances} 
as estimates of the final success rates for our desired $d\fin$. 
For the protocols that implement escape via stabilizer measurements, we assume three rounds for $\fd3$ and five rounds for $\fd5$.

The circuit depth of each segment in different protocols depends on the specific implementation of gates and may be reduced through circuit optimization. 
To estimate the spacetime volumes $V_i$ for the protocols in Ref.~\cite{gidney2024magic, chen2025efficient}, we use their reported Stim circuits and count the circuit depth without any further optimization. 
For our protocol, we perform the same analysis using the circuits provided in~\cref{app:msc-circuits}. 
The number of active qubits and the circuit depth of each segment for different {$\mathsf f=3$ and $\mathsf f=5$} protocols are summarized in~\cref{table: STV fd3 summary} and~\cref{table: STV fd5 summary}, respectively.

\subsection{Shot survival}

\begin{figure*}
    \centering
    \begin{tikzpicture}
        \node at (0, 0) {
            \includegraphics[
                height = 0.32\linewidth,
                trim = 0 10 0 10, 
                clip
                ]{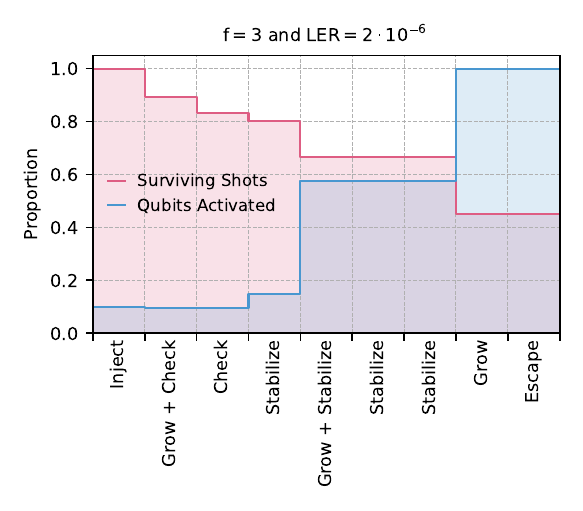}
            \subfloat{\label{fig:appe_fd3_staircase}}%
            \includegraphics[
                height = 0.32\linewidth,
                trim = 0 10 0 10, 
                clip
                ]{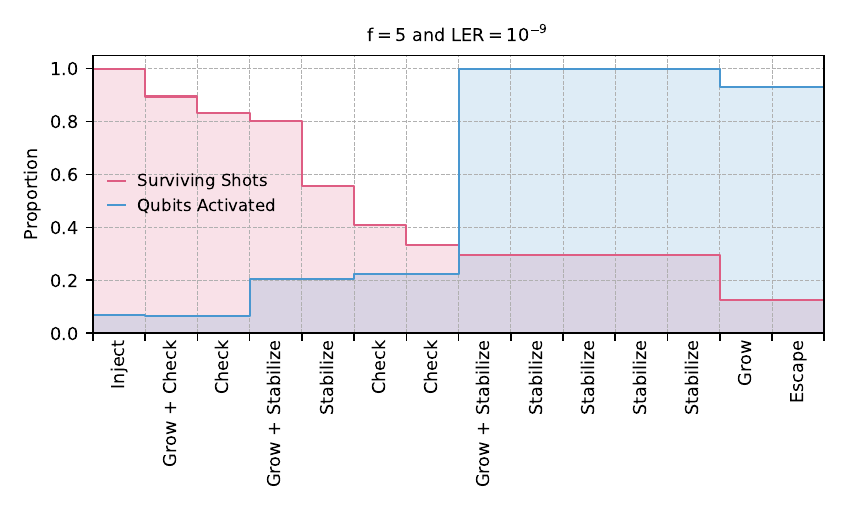}
            \subfloat{\label{fig:appe_fd5_staircase}}%
                };
        \node at (-7.5, 2.7) {\textbf{\textsf{(a)}}};
        \node at (-0.25, 2.7) {\textbf{\textsf{(b)}}};
    \end{tikzpicture}
    \caption{
        {
        \textbf{Survival of shots during cultivation.}
        Fraction of activated qubits (orange) and survival rate of shots (blue) throughout the protocol.
        We calculate this for a target logical error rate of (a) $2\times10^{-6}$ with $\fd3$, and (b) $10^{-9}$ with $\fd5$.
        This includes gap-based postselection and unitary growth during the escape stage.}
        }
    \label{fig:appe_staircases}
\end{figure*}

\Cref{fig:appe_staircases} shows the `survival' of shots for the complete cultivation process, including gap-based postselection.
We do this for a target logical error rate of $2\times10^{-6}$ ($\fd3$) and $10^{-9}$ ($\fd5$). 
The cultivation procedure is divided into multiple stages, each corresponding to one round of qubit measurements, following the convention in Ref.~\cite{gidney2024magic}. 
In our plots, we assume that the final unitary growth is applied only when the shot passes the gap-postselection. 
The orange curve indicates the fraction of qubits activated in each stage of the protocol, while the blue curve shows the survival rate of shots as the process progresses. 
The final acceptance rate is determined by the survival rate in the last stage, where the code has successfully passed all postselection checks, and has reached the target logical error rate.

\begin{table*}
\centering
    \begin{tabular}{|c|l|l|l|}
    \hline
    \textbf{Protocol} & \textbf{Stage} & \textbf{Spacetime volume} & \textbf{Description} \\ \hline
    \makecell[c]{This \\ work} &
    \makecell[l]{Inject \\
                Grow \\
                Check $\times2$ \\
                Stabilize \\
                Grow \\
                Stabilize $\times3$ \\
                Grow
                } &
    \makecell[l]{$9\cdot 6 + 17\cdot 8$ \\
                $13\cdot 4$ \\
                $2\times [2(3\cdot 4)+16\cdot 3]$ \\
                $25\cdot 8$ \\
                $25\cdot 4$ \\
                $3\times (97\cdot 8)$ \\
                $169\cdot 4$
                } &
    \makecell[l]{Unitary encoding of $\rot3$. \\
                Unitary growth to $\reg3$. \\
                $\bar{H}\XY$ check; includes encoding/decoding $\GHZ{3}$. \\ 
                Stabilizer measurements of $\reg3$. \\ 
                Unitary growth to $\rot5$. \\ 
                Stabilizer-based growth to $\rot7$.  \\
                Unitary growth to $\rot{13}$.
                } \\ \hline
   ~\cite{gidney2024magic} & 
    \makecell[l]{Inject \\
                Stabilize \\
                Check \\
                Stabilize \\
                Escape \& stabilize
                } &
    \makecell[l]{$13\cdot 15$ \\
                $14\cdot 10$ \\
                $2\times (13\cdot 7)$ \\
                $3\times (337 \cdot 12)$ \\
                $3\times (337 \cdot 12)$
                } &
    Stages defined as in Fig.~15(left) of Ref.~\cite{gidney2024magic}.\\ \hline       
    
   \cite{chen2025efficient} & 
    \makecell[l]{Inject \\
                Stabilize \\
                Morph \\
                Check \\
                Morph \& expand \\
                Stabilize} &
    \makecell[l]{$9\cdot 3 + 19 \cdot 9$ \\
                $19\cdot 9$ \\
                $15\cdot 5$ \\
                $2\times (24 \cdot 9)$ \\
                $343\cdot 10$ \\
                $3\times (337 \cdot 8)$} &
    Stages defined as in Fig.~14(a) of Ref.~\cite{chen2025efficient}.\\ \hline   
    \cite{claes2025cultivatingtstatessurface} & 
    \makecell[l]{Inject \\
                Stabilize \\
                Deform \\
                Check \\
                Undeform \\
                Stabilize} &
    \makecell[l]{$9\cdot 10$ \\
                $17 \cdot 8$ \\
                $21 \cdot 10$ \\
                $2\times (26 \cdot 12)$\\
                $26 \cdot 10$ \\
                $3\times (337 \cdot 8)$} &
    Stages defined as in Fig.~1(a) of Ref.~\cite{claes2025cultivatingtstatessurface}.\\ \hline  
    \cite{vaknin2025magic} & 
    \makecell[l]{Stabilize \\
                Inject \& stabilize \\
                Grow \\
                Check \\
                Grow \\
                Escape \& stabilize} &
    \makecell[l]{$17 \cdot 8$ \\
                $17 \cdot 17$ \\
                $17 \cdot 4$ \\
                $2\times (24 \cdot 9)$ \\
                $17 \cdot 4$ \\
                $3\times (337 \cdot 8)$} &
    Stages defined as in Ref.~\cite{vaknin2025magic}.\\ \hline  
    \end{tabular}
    \caption{\textbf{Spacetime volume comparison for $\boldsymbol{\fd3}$} for different protocols. 
    Each protocol is divided into multiple stages with their corresponding overhead. 
    Dot-product notation indicates qubit $\cdot$ steps, while expressions with an explicit `$\times$' indicate repeated execution of the circuit. 
    For example, $A\times(Q\cdot T)$ denotes a circuit with $Q$ qubits and $T$ time steps repeated $A$ times. 
    Note that in all circuits, the preparation of a $\ket{+}$ state is implemented as preparation of $\ket{0}$ followed by a Hadamard gate, taking two steps. 
    Similarly, each $X$-basis measurement is counted as two steps.}
    \label{table: STV fd3 summary}
\end{table*}

\begin{table*}
\centering
    \begin{tabular}{|c|l|l|l|}
    \hline
    \textbf{Protocol} & \textbf{Stage} & \textbf{Spacetime volume} & \textbf{Description} \\ \hline
    \makecell[c]{This \\ work} &
    \makecell[l]{Inject \\ 
                Grow \\
                Check $\times2$ \\
                Grow \\
                Stabilize $\times2$ \\
                Grow \\
                Check $\times2$ \\
                Grow \\
                Stabilize $\times5$ \\
                Grow
                } &
    \makecell[l]{$9\cdot 6 + 17\cdot 8$ \\
                $13\cdot 4$ \\
                $2\times [2(3\cdot 4)+16\cdot 3]$ \\
                $25\cdot 4$\\
                $2\times (49 \cdot 8)$ \\
                $49 \cdot 4$ \\
                $2\times (6\cdot 8 + 54\cdot 5 + 5\cdot 5)$ \\
                $81\cdot4$ \\
                $5\times (241\cdot8)$ \\
                $225\cdot4$
                } &
    \makecell[l]{Unitary encoding of $\rot3$. \\
                Unitary growth to $\reg3$. \\
                $\bar{H}\XY$ check; includes encoding/decoding $\GHZ{3}$. \\ 
                Unitary growth to $\rot5$. \\ 
                Stabilizer measurements of $\rot5$. \\ 
                Unitary growth to $\reg5$. \\ 
                $\bar{H}\XY$ check; includes encoding/decoding $\GHZ{5}$. \\
                Unitary growth to $\rot9$. \\
                Stabilizer-based growth to $\rot{11}$. \\
                Unitary growth* to $\rot{15}$.
                } \\ \hline
                
   ~\cite{gidney2024magic} & 
    \makecell[l]{Inject \\
                Stabilize \\
                Check \\
                Stabilize \\
                Check \\
                Stabilize \\
                Escape \& stabilize
                } &
    \makecell[l]{$13\cdot 15$ \\
                $14\cdot 10$ \\
                $2\times(13\cdot 7)$ \\
                $3\times (38 \cdot 12)$ \\
                $2\times(39\cdot 12)$ \\
                $5\times (449\cdot12)$ \\
                $5\times (449\cdot12)$
                } &
    Stages defined as in Fig.~15(left) of Ref.~\cite{gidney2024magic}.\\ \hline       
    
   ~\cite{chen2025efficient} & 
    \makecell[l]{Inject \\
                Stabilize \\
                Morph \\
                Check \\
                Morph \& grow \\
                Stabilize \\
                Morph \\
                Check \\
                Morph \& expand \\
                Stabilize
                } &
    \makecell[l]{$9\cdot 3 + 19 \cdot 9$ \\
                $19\cdot 9$ \\
                $15\cdot 5$ \\
                $2\times (24 \cdot 9)$ \\
                $57 \cdot 14$ \\
                $51 \cdot 9$ \\
                $35 \cdot 6$ \\
                $2\times (50 \cdot 12)$ \\
                $250 \cdot 14$ \\
                $5 \times (449 \cdot 8)$
                } &
    Stages defined as in Fig.~14(a) of Ref.~\cite{chen2025efficient}.\\ \hline  
    
    ~\cite{claes2025cultivatingtstatessurface} & 
    \makecell[l]{Inject \\
                Stabilize \\
                Deform \\
                Check \\
                Undeform \\
                Expand \& stabilize \\
                Deform \\
                Check \\
                Undeform \\
                Stabilize
                } &
    \makecell[l]{$9\cdot 10$ \\
                $17 \cdot 8$ \\
                $21 \cdot 10$ \\
                $2\times (26 \cdot 12)$\\
                $26 \cdot 10$ \\
                $25 \cdot 4 + 49 \cdot 8$ \\
                $57 \cdot 10$ \\
                $2\times (66 \cdot 17)$ \\
                $66\cdot 10$ \\
                $5\times (449 \cdot 8)$
                } &
    Stages defined as in Fig.~1(a) of Ref.~\cite{claes2025cultivatingtstatessurface}.\\ \hline  
    
    \cite{vaknin2025magic} & 
    \makecell[l]{Stabilize \\
                Inject \& stabilize \\
                Grow \\
                Check \\
                Grow \\
                Grow \\
                Stabilize \\
                Grow \\
                Check \\
                Grow \\
                Grow \& stabilize
                } &
    \makecell[l]{$17 \cdot 8$ \\
                $17 \cdot 17$ \\
                $17 \cdot 4$ \\
                $2\times (24 \cdot 9)$ \\
                $17 \cdot 4$ \\
                $25 \cdot 13$ \\
                $49 \cdot 8$ \\
                $49 \cdot 4$ \\
                $2\times (62 \cdot 11)$ \\
                $49 \cdot 4$ \\
                $5\times (449 \cdot 8)$
                } &
    Stages defined as in Ref.~\cite{vaknin2025magic}.\\ \hline  
    \end{tabular}
    \caption{
    {\textbf{Spacetime volume comparison for $\boldsymbol{\fd5}$} for different protocols.} 
    Each protocol is divided into multiple stages with their corresponding overhead. 
    Notation identical to that of~\cref{table: STV fd3 summary}.
    (*Note that this stage uses the unitary growth circuit from~Ref.~\cite{claes2025lowerdepthlocalencodingcircuits}.)}
    \label{table: STV fd5 summary}
\end{table*}

\end{document}